\documentclass[12pt]{article}
\usepackage{geometry}
\usepackage{amsmath,amssymb}
\usepackage[nosort]{cite}
\usepackage{float}
\usepackage{latexsym}
\usepackage{graphicx}
\usepackage{color}
\usepackage{subfig}

\newcommand{\fft}[2]{{\frac{#1}{#2}}}

\newcommand{\beq}{\begin{equation}}
\newcommand{\be}{\begin{equation}}
\newcommand{\ee}{\end{equation}}
\newcommand{\bea}{\begin{eqnarray}}
\newcommand{\eea}{\end{eqnarray}}
\renewcommand{\Re}{\mathrm{Re}\,}

\newcommand{\nn}{\nonumber}

\newcommand{\ph}{\hat{\phi}}
\newcommand{\pa}{\partial}
\newcommand{\pau}{\partial_u}
\newcommand{\pax}{\partial_{x_4}}
\newcommand{\paz}{\partial_{z}}
\newcommand{\hF}{\hat{F}}
\newcommand{\pahu}{\partial_{\hat{u}}}
\newcommand{\hu}{\hat{u}}
\newcommand{\hhp}{\tilde{\phi}}
\newcommand{\pap}{\partial_{\rho}}
\newcommand{\paq}{\partial_{q}}
\makeatletter \@addtoreset{equation}{section} \makeatother

\begin{document}

\begin{titlepage}
\hbox to \hsize{\hspace*{0 cm}\hbox{\tt }\hss
    \hbox{\small{\tt TAUP-2861/07}}}
\hbox to \hsize{\hbox{ }\hss
    \hbox{\small{\tt UTTG-04-07}}}

\vspace{1 cm}

\centerline{\bf \Large Localized Backreacted Flavor Branes }

\vspace{.6cm}

\centerline{\bf \Large in Holographic QCD}

%\vspace{.6cm} \centerline{\bf \Large }
\vspace{1 cm}
 \centerline{\large $^\dagger\!\!$ Benjamin A. Burrington, ${\;^*\!\!}$ Vadim S. Kaplunovsky,}
  \centerline{\large $\;^{\dagger\!\!}$ Jacob Sonnenschein}

\vspace{.5 cm}
\centerline{$^{\dagger}${\it School of Physics and Astronomy,}}
\centerline{\it The Raymond and Beverly Sackler Faculty of Exact Sciences,}
\centerline{\it Tel Aviv University, Ramat Aviv, 69978, Israel.}

\vspace{.5 cm}

\centerline{$^*${\it University of Texas at Austin}}
\centerline{\it Theory Group, Physics Department,}
\centerline{\it 1 University Station, C1608}
\centerline{\it Austin, TX 78712–1608, USA}

\vspace{0.3 cm}

\begin{abstract}
We investigate the perturbative (in $g_s N_{D8}$) backreaction of
localized D8 branes in D4-D8 systems including in particular the
Sakai Sugimoto model. We write down the explicit expressions of the
backreacted metric, dilaton and RR form.
 We
find that the backreaction remains small up to a radial value of $u
\ll \ell_s/(g_s N_{D8})$, and that the background functions are
smooth except at the D8 sources.  In this perturbative window, the
original embedding remains a solution to the equations of motion.
Furthermore, the  fluctuations around the original embedding,
describing scalar mesons, do not become tachyonic due to the
backreaction in the  perturbative regime. This is is due to a
cancelation between the DBI and CS parts of the D8 brane action in
the perturbed background.
\end{abstract}
\end{titlepage}

\tableofcontents

%%%%%%%%%%%%%%%%%%%%%%%%%%%%%%%%%%%%%%%%%%
\section{Introduction}
%%%%%%%%%%%%%%%%%%%%%%%%%%%%%%%%%%%%%%%%%%

By now, AdS/CFT has become a standard tool in theoretical physics for the study of gauge theories at strong coupling.
In many ``stringy'' models of gauge dynamics fundamental matter is included by embedding a set of ``flavor branes'' in addition to the ``glue/color branes.''  In such a setup, the strings connecting only to the ``glue branes'' are in the adjoint of the $U(N_c)$ group, giving gauge particles (multiplets), and those connecting only to the ``flavor branes'' are in the adjoint of $U(N_f)$, giving the mesons (meson multiplets), and those connected to both the ``flavor'' and ``glue'' branes are in the fundamental representation of both the $U(N_c)$ and $U(N_f)$ groups, giving the quarks (matter/quark multiplets). Anti-quarks are obviously depicted by similar strings with the opposite orientation.

In principle for large $N_c$ and large $N_f$ one could go to a
combined near horizon limit, translate the branes into fluxes
\footnote{bearing in mind that one must keep the open string
spectrum associated with the flavor branes, even though they are
producing macroscopic flux.  Such a case is a full backreaction, but
not a decoupling limit.  Recall that global symmetries of the field
theory translate into gauge symmetries in the gravity: gauging a
large $N_f$ flavor group using supergravity alone is unfeasible.}
and derive the gravity background that is a holographic dual of a
gauge system with gluons and quarks ( and, if the model is
supersymmetric, their supersymmetric partners).  However, in
practice such models are being constructed using the probe
approximation.  In this approximation one uses a gravity
background built from the near horizon limit of a large $N_c$ glue
branes and adds to it a set of $N_f$  flavor probe branes. The basic
assumption  of the probe approximation is that for the case of
 $N_f \ll N_c$  the backreaction of the probes on the background can be neglected. The flavor physics is then extracted by analyzing the effective action that describes the flavor branes in the glue background, namely the DBI action plus the CS action. This practice was introduced
in \cite{Karch} in the context of the $AdS_5\times S^5$ model, for a confining background in
\cite{SakSon} and subsequently in a large variety of other models\cite{Babington:2003vm},\cite{KMMW2}, \cite{problist}.
Exceptions to this are certain  fully
backreacted non critical models like
\cite{Klebanov:2004ya},\cite{Bigazzi:2005md},\cite{Casero:2005se},
and \cite{Gursoy:2007er}.

A landmark holographic model of chiral symmetry and chiral symmetry
breaking is the model of Sakai and Sugimoto\cite{Sakai:2004cn}. This
model is based on the incorporation of  a stack of $N_f$ D8 and
$N_f$ anti- D8  flavor branes into the background of near extremal
$D4$ branes\cite{Witten:1998zw}. In the latter background one
compactifies one of the world volume coordinates of the D4 branes on
a circle of radius $R$. For energies $E\ll 1/R$ the background
describes a four dimensional system with gluon degrees of freedom
plus contaminating Kaluza Klein modes. The profile of the flavor
branes determined by the DBI action is that of a U shape. This
provides a simple geometrical picture of chiral symmetry breaking,
namely, for large radial direction (see figure \ref{D8-barD8SS}),
which corresponds to the UV limit of the gauge theory,  the stack of
the D8 branes and of the anti-D8 branes are separated and hence
there is a $U_L(N_f)\times U_R(N_f)$ symmetry, and in the IR limit
the two stacks merge one into the other and thus only the diagonal
$U(N_f)$ survives as a symmetry. A variety of physical properties of
meson and baryon physics has been extracted from the model. These
include the massive meson spectrum, the massless Goldstone
pions\cite{Sakai:2004cn}, certain decay rates \cite{Sakai:2005yt}
 as well as the thermal behavior of hadrons \cite{Aharony:2006da},\cite{Peeters:2006iu}.

The validity of the model \cite{Sakai:2004cn}, is the same as all
other probe models: $N_f \ll N_c$. To contact to  real hadron
physics, one obviously is interested in the case where the number of
flavors is similar to that of the colors and both are not large. To
get down to small $N_c$ one will have to invoke a full string theory
rather than a effective gravity model. However, increasing  the
ratio of $N_f/N_c$  can still be done in the context of an effective
field theory, provided we go beyond the probe approximation and
incorporate the backreaction of the flavor branes on the gravity
background.  This may enable us to determine the flavor dependence
of certain physical properties of the gauge theory which we expect
to be $N_f$ dependent, for example the beta function, or the ratio
of viscosity to entropy density of the quark-gluon fluid
\cite{Kovtun:2004de}.\footnote{In \cite{Mateos:2006yd}, the leading
order correction in $N_f/N_c$ of this property was determined in the
context of a model with D7 branes in near extremal $Ads_5\times S^5$
background.  However, in this model it was shown that the zero mode,
which is equivalent to smearing, is all that is necessary to this
level in $N_f/N_c$.}

Similar studies for localized backreactions in D3-D7 systems include \cite{Kehagias:1998gn,Aharony:1998xz,Grana:2001xn,Bertolini:2001qa,Burrington:2004id}.  In \cite{Kehagias:1998gn}, very general framework for studying type IIB supergravity with metric/five form and holomorphic dilaton/axion.
The work of \cite{Aharony:1998xz,Grana:2001xn,Bertolini:2001qa} studied when the D7 branes are located at singular points in manifolds, and \cite{Burrington:2004id} studied the form of the solution to the equations following from the D3-D7 system and effects on other probe branes in such backgrounds.  All these studies worked using the supergravity alone, while here we will derive delta function source terms from an action of the form
\be
S_{Bulk} +K_8(S_{DBI}+S_{CS}). \label{actionform}
\ee
We will use this action to determine how to source the bulk fields.  Although we obtain the full equations of motion from this, we will study their solutions in a perturbative limit.  Therefore, while we are going beyond the probe approximation, we are still confined to the regime $N_f \ll N_c$ for the simple reason that we want the series in powers of $N_f/N_c$ to converge quickly.

In fact there is an even more important motivation to explore the
model of \cite{Sakai:2004cn} beyond the probe approximation, and
that is the issue of the stability of the model.  One may wonder
whether the model is stable only in the probe approximation and that
the backreaction of the probe branes on the background does not
destabilize the setup.   A simplified picture of the model is that
of the circle of the compactified direction  with the two endpoints
of the stacks of probe branes and anti branes which can be
represented as a $+N_f$ charge located at one point on the circle
and $-N_f$ charge located at the antipodal point. In this simplified
``electrostatic'' setup if one of the charges gets a slight
perturbation in one direction it will be attracted to the opposite
charge and will not be driven back to its original location.
Moreover, the antipodal setup described in   \cite{Sakai:2004cn} has
been generalized to a family of setups where the separation distance
between the brane and anti-brane is taken to be $L\leq \pi R$. For
these cases the ``electrostatic instability'' is even more severe.
The question is therefore whether this naive intuition is justified
and the backreaction of the probe branes indeed destabilizes the
model. On the other hand there is a naive argument why the
perturbative backreacted system should be stable and non tachyonic.
Since  the gauge holographic dual of the model before purturbing it
has a spectrum with a mass gap ( apart from the pions), a small a
small perturbation cannot bridge the gap and produce tachyonic
modes\cite{Aharony:2002vp}.

Further, one may wonder what happens to the the dilaton tadpole condition, given that this is a D8 $\bar{\rm D8}$ on a circle, and both branes and anti branes source the dilaton in the same way.
Hence, for these codimension one flavor branes  one anticipates that the dilaton will have a cusp behavior at the location of the probe branes as well as a cusp (and not an anti-cusp) at the anti-brane. It seems naively that there is no way to sew together these two cusps.

Thus the goal of this paper is to compute the leading order
backreacted background, and address the stability, and the dilaton tadpole.
To do so, we write down the full action of the system (of the form in (\ref{actionform}))
which is composed of the action of  massive type $II_A$ supergravity
and the nine dimensional DBI +CS actions associated with the the
$D8$ flavor branes. At this point one often invokes a smearing of
the flavor branes along their transverse direction
\cite{Benini:2006hh,Benini:2007gx,Arean:2007nh} which renders the
combined action into a ten dimensional one. This approach simplifies
the analysis by turning the equations of motion (EOMs) into ordinary
differential equations (ODEs) of some radial variable. However, we
expect that certain physical questions may not be answered using
this procedure, for example the stability discussed above. Thus we
avoid using the smearing technique and we keep the flavor branes as
localized objects. This yields delta function source terms for the
equations of motion of the graviton, dilaton and the $F_{(10)}$ RR
field strength form  associated with the D8 branes. The complexity
of these equations is increased, relative to the smearing technique,
because the EOMs are now partial differential equations (PDEs); the
relevant functions must depend on the coordinate(s) transverse to
the flavor brane. \footnote{For this added complication, though, we
simplify the equations by using a perturbative approach.  In some
sense, this is complimentary to smearing: one smears the branes to
obtain non-linear ODEs to solve; we instead perturb the equations to
obtain linear PDEs.}

We solve these equations perturbatively to the leading order in $N_f/N_c$.  We take 3 cases for the background to help address the questions in stages, and gain intuition for how the solutions should behave.

\begin{figure}[t]
\begin{center}
\vspace{3ex}
\includegraphics[width=0.8\textwidth]{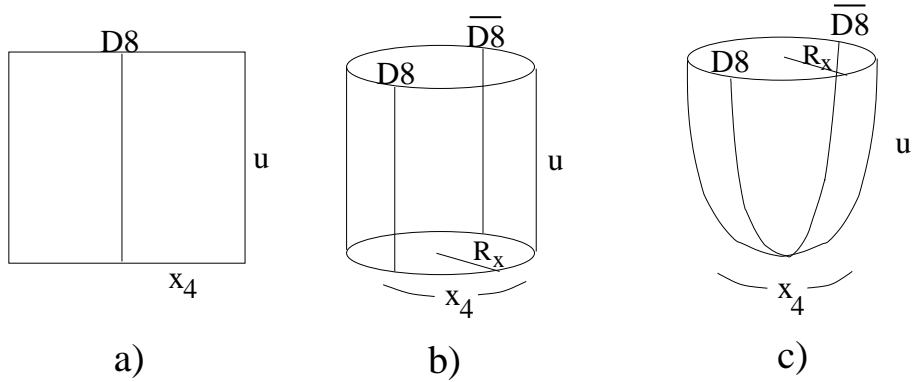}\end{center}
\caption{The three cases studied: a) the uncompactified $x_4$ case, b) the compactified $x_4$ case, and c) the near extremal ``cigar'' case. \label{3cases}}
\end{figure}

We first solve for the simplified system of a decompacitfied transverse coordinate of the D8 branes, which has been studied in its own right in  \cite{Antonyan:2006vw}, \cite{Davis:2007ka}(see figure \ref{3cases} (a)). For this case we were able to find exact solutions of the partial differential equations.
To our surprise we have found that whereas the solutions for the perturbations of some background fields behave, as we have expected, with a cusp at the location of the probe branes (a ``$\Lambda$'' shape),
for other the behavior is of an inverted cusp dressed with a double hump structure (an ``m'', see figure \ref{graphq}).
We explore the ranges where we expect the supergravity to be a good description and find that $u\gg \ell_s^4/R_{D4}^3$, and that the perturbative results are good up to $u\ll 1/Q_f\equiv 4\pi \ell_s/(g_s N_f)$.  Further, from this study, it becomes plausible that compactifying the $x_4$ direction is possible, as the functions die off at large $x_4$.

Next, we address the compactified (but extremal) case both to view the effects of compactification, but also as a rough approximation of the ``cigar'' case at large $u$ (see figure \ref{3cases} (b)). We first treat this case where we sum over the images of the uncomactified case, and then as a Fourier decomposition.  From this we see that there is no issue with the dilaton tadpole constraint: the two cusps meets smoothly.  We find that the decompactified limit emerges at large $u$.  Both methods are applied because, although both series always converge, one series converges very quickly at large $u$, while the other converges quickly at small $u$.

Finally we analyze the system of the near extremal D4 branes.  We find that in this case the perturbation theory is good for $u\ll 1/Q_f$ which is generically stronger than the $u^3\ll R_{D4}^3/g_s^{4}$ supergravity regime.  Further, we find that the supergravity description is valid near $u=U_K$, with the additional constraint $(U_K/R_{D4})^{3/4}Q_f\ll 1/\ell_s$.  This translates into the requirement that $(T_{st}/M_{gb}^2)^{1/2}\lambda_4 N_f/N_c\ll1$ where $T_{st}$ is the string tension, $M_{gb}$ is the typical glueball mass, and $\lambda_4=g_4^2 N_c$ is the 't Hooft coupling. For large values of the radial direction the solution is obviously like that of the extremal compactified case. We use a Fourier decomposition to show that a finite expansion around the tip of the cigar is possible, and then implement this expansion for the first few terms.

Once we have established the perturbative solutions, we proceed to analyze the stability of the system. We first show that the solution of the embedding of the flavor brane at the probe level persists also in the leading order backreaction.  We further show that the fluctuations of the embedding, which correspond to scalar mesons in the dual gauge theory, are non tachyonic. Hence we shown that the system is stable at least for an action that is quadratic in the fluctuations.  This is due to a cancelation between the electrostatic repulsion (CS action) and the gravitational attraction (DBI action).  Hence, the above analogy with an electrostatic problem is not quite justified: the electric repulsion is canceled by a gravitational attraction.  The only other force is that of the tension of the brane, which is restorative.  The corrections to this force, while interesting, cannot change the qualitative feature of stability while the perturbative analysis is valid (however, the effects of the non-perturbative backreaction is still an open question).

The paper is organized as follows. In the next section (2) we briefly review the general setup of the problem, namely, the Sakai Sugimoto \cite{Sakai:2004cn} model and the massive type $II_A$ supergravity action \cite{Romans:1985tz,Bergshoeff:1995vh}. In section 3  we write the supergravity EOMs that incorporate the backreaction of the probe branes. We then introduce an ansatz for the metric which we substitute into the equations.  The perturbative parameter is defined, and these equations are expanded. The gauge invariance, in the form of small coordinate transformations, of the system is discussed. In section 4  we present the solutions of the backreacted EOM. We start with the solutions for the uncompactified case, and then by summing over images the solutions for the compactified extremal case is constructed.  We also use Fourier analysis to study this case. This enables us to determine the UV behavior of the near extremal case because the geometries are identical at large radius. The third step is to write down the solutions for the near extremal case in the region close to the horizon. In the following section we analyze the stability of the system. We first show that the solution of the EOMs that follow from the backreacted DBI+CS action are the same as those of the unperturbed solution. Finally we shown that the spectrum of fluctuations around this embedding is tachyon free and hence we conclude that to the leading order in $N_f/N_c$ the system is stable.

\section{ Brief review of the general setup}
Before we start the analysis of the backreaction of the flavor branes, we briefly review the two main ingredients of the
general setup of the problem, namely, the Sakai Sugimoto model and the action of the massive type $II_A$ supergravity.  The reader familiar with these topics should skip to the next section.
%%%%%%%%%%%%%%%%%%%%%%%%%%%%%%%%%%%%%%%%%%
\subsection{Sakai-Sugimoto (SS) model}
%%%%%%%%%%%%%%%%%%%%%%%%%%%%%%%%%%%%%%%%%%

Constructing holographic models duals of gauge dynamics that admits confinement is by now a relatively easy task.
Incorporating flavor chiral symmetry, on the other hand, turns out to be more complicated. A prototype model that
includes both phenomena is the
 Sakai-Sugimoto model \cite{Sakai:2004cn}. It  is a  model
of a holographic dual for a $3+1$ dimensional gauge theory with a
continuous $SU(N_f)\times SU(N_f)$ flavor chiral symmetry which is
spontaneously broken. It is based on the incorporation of  $N_f$ D8-branes
 and $N_f$ anti-D8-branes into Witten's model \cite{Witten:1998zw}. The latter describes the near horizon limit of $N_c$ D4-branes, compactified on a circle of radius $R$ ($x_4
\equiv x_4 + 2 \pi R_x$) with anti-periodic boundary conditions for the
fermions. The  D8-branes are placed at
$x_4=0$ and the anti-D8-branes at $x_4=L$. The gauge theory dual of this SUGRA setup is
a $4+1$ dimensional $SU(N_c)$ maximally supersymmetric
gauge theory, compactified on a circle with anti-periodic boundary
conditions for the adjoint fermions, and coupled to $N_f$ left-handed
fermions in the fundamental representation of $SU(N_c)$ localized at
$x_4=0$, and to $N_f$ right-handed fermions in the fundamental
representation localized at $x_4=L$.

The  basic assumption of the model is that in the limit of $N_f \ll
N_c$ one can ignore the back-reaction of the $N_f$ D8 branes and
$N_f$ anti-D8 branes. As mentioned above the  goal of the present
work is to examine in details the back-reaction of the D8 and
anti-D8 on the background. With the probe assumption the closed type
IIA string background is given by :
\begin{equation}
\label{SSmodel}
\begin{aligned}
ds^2 &= \left(\frac{u}{R_{D4}}\right)^{3/2} \left[-dt^2 + \delta_{ij}
dx^i dx^j + f(u) dx_4^2\right] + \left(\frac{R_{D4}}{u}\right)^{3/2}
\left[\frac{du^2}{f(u)}+u^2
d\Omega_4^2\right]\,,\\
F_{(4)} &= \frac{3 R_{D4}^3}{g_s} \Omega_4,\quad e^{\phi} = g_s
\left(\frac{u}{R_{D4}}\right)^{3/4},\quad R_{D4}^3 \equiv \pi g_s
N_c \, l_s^3, \quad f(u) \equiv 1 -
\left(\frac{U_{K}}{u}\right)^3\,,
\end{aligned}
\end{equation}
where $t$ is the time direction and $x^i$ ($i=1,2,3$) are the
 uncompactified world-volume coordinates of the D4 branes, $x_4$ is a
 compactified direction of the D4-brane world-volume which is transverse to
 the probe D8 branes,
 $d\Omega_4^2$ is the metric of a unit four-sphere and
$\epsilon_4$ is its volume form, and $g_s$ is related to the $4+1$
dimensional gauge coupling by $g_5^2 = (2\pi)^2 g_s l_s$. The
submanifold spanned by $x_4$ and $u$ has the topology of a cigar
with $u \geq U_{K}$, and requiring that this has a non-singular
geometry gives a relation between $U_{K}$ and $R_x$,
\begin{equation}
R_x = \frac{2}{3} \left(\frac{R_{D4}^3}{U_{K}}\right)^{1/2}.
\end{equation}

The parameters of this gauge theory, the five-dimensional gauge coupling
$g_5$, the low-energy four-dimensional
 gauge coupling $g_4$, the glueball mass scale $M_{gb}$, and the
 string tension $T_{st}$ are determined from the background
 (\ref{SSmodel}) in
 the following form  :
\bea\label{stringauge}
g_5^2&=&(2\pi)^2 g_s l_s,\qquad
g^2_{4}=\frac{g_5^2}{2\pi R_x}=
3\sqrt{\pi}\left ( \frac{g_s U_K}{N_c l_s}\right )^{1/2},
 \qquad
M_{gb} = \frac{1}{R_x},
%=\frac{3}{2\sqrt{\pi}}\frac{\sqrt{u_\Lambda}}{\sqrt{g_s N_c
% l_s^3}}
\nonumber \\ \cr  T_{st} &=& \frac{1}{2\pi
 l_s^2}\sqrt{g_{tt}g_{xx}}|_{u=U_K}= \frac{1}{2\pi l_s^2}\left(
 \frac{U_K}{R_{D4}} \right)^{3/2} =\frac{2}{27\pi} \frac{g^2_4 N_c}{R_x^2}
= \frac{\lambda_5}{27\pi^2 R_x^3},\eea
where $\lambda_5 \equiv g_5^2 N_c$, $M_{gb}$ is the typical scale of
the glueball masses computed from the spectrum of excitations around
(\ref{SSmodel}), and $T_{st}$ is the confining string tension in this
model (given by the tension of a fundamental string stretched at $u=u_{K}$
where its energy is minimized). The gravity approximation is valid
whenever $\lambda_5 \gg R_x$, otherwise the curvature at $u \sim
U_{K}$ becomes large. Note that as usual in gravity
approximations of confining gauge theories, the string tension is much
larger than the glueball mass scale in this limit. At very large values of $u$ the
dilaton becomes large, but this happens at values which are of order
$N_c^{4/3}$ (in the
large $N_c$ limit with fixed $\lambda_5$), so this will
play no role in the large $N_c$ limit that we will be interested in.
The Wilson line of this gauge theory (before putting in the D8-branes)
admits an area law behavior \cite{BISY2}, as can be easily seen using
the conditions for confinement of \cite{KSS}.

The gauge theory dual to the SUGRA background (\ref{SSmodel}) is in fact not
 four dimensional even
  at energies lower than the Kaluza-Klein scale $1 / R_x$  since
   the masses of the glueballs are also  $M_{gb}=1/R_x$, namely,   there
is no real separation between the confined four-dimensional fields and
the higher Kaluza-Klein modes on the circle.
As discussed in \cite{Witten:1998zw},
in the opposite limit of $\lambda_5 \ll R_x$, the theory approaches the
$3+1$ dimensional pure Yang-Mills theory at energies small compared to
$1/R_x$, since in this limit the scale of the mass gap is exponentially
small compared to $1/R$.

The probe flavor D8-branes
span the coordinates $t, x^i, \Omega_4$, and trace  a  curve $u(x_4)$
in the $(x_4,u)$-plane. Near the boundary at $u\to \infty$ we want to have
$N_f$ D8-branes localized at $x_4=0$ and $N_f$ anti-D8-branes (or D8-branes
with an opposite orientation) localized at $x_4=L$. Naively one might think
that the D8-branes and anti-D8-branes would go into the interior of the space
and stay disconnected; however, these 8-branes
do not have anywhere to end in the background (\ref{SSmodel}), so the
form of $u(x_4)$ must be such that the D8-branes smoothly connect to the
anti-D8-branes (namely, $u$ must go to infinity at $x_4=0$ and at $x_4=L$,
and $du/dx_4$ must vanish at some minimal $u$ coordinate $u=u_0$). Such
a configuration spontaneously
breaks the chiral symmetry from the symmetry group which is
visible at large $u$,
$U(N_f)_L\times U(N_f)_R$, to the diagonal $U(N_f)$ symmetry.
Thus, in this configuration the topology forces a breaking of the chiral
symmetry.

\begin{figure}[t]
\begin{center}
\vspace{3ex}
\includegraphics[width= 80mm]{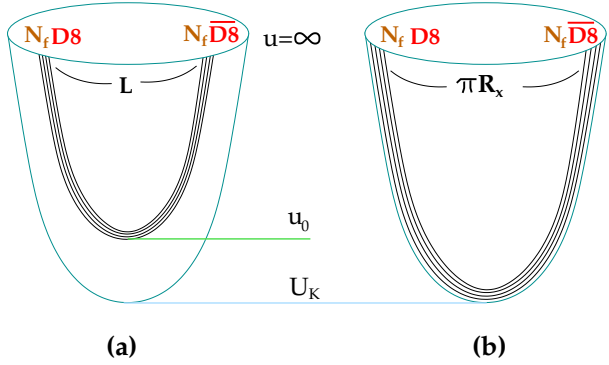}\end{center}
\caption{The dominant configurations of the D8 and anti-D8 probe
branes in the Sakai-Sugimoto model at zero temperature,
which break the chiral symmetry. The same configurations will turn
out to be relevant also at low temperatures.
On the left a generic configuration with an asymptotic separation of
$L$, that stretches down to a minimum at $u=u_0$, is drawn. The figure
on the right describes the limiting antipodal case $L=\pi R_x$, where
the branes connect at $u_0=U_{K}$. \label{D8-barD8SS}}
\end{figure}

To determine the profile of flavor probe branes, one has to solve the equations of motion of
that follow from the DBI + CS action that describes the probe branes.
It is easy to check that the CS term in the D8-brane action does not affect the
solution of the equations of motion.  More precisely, the equation
of motion of the gauge field has a classical solution of a vanishing
gauge field, since the CS term includes terms of the form $C_5\wedge
F\wedge F$ and $C_3 \wedge F\wedge F\wedge F$. So, we are left only with
the DBI action.
The induced metric on the D8-branes  is
\bea
\label{induced}
ds^2_{D8}&=&\left( \frac{u}{R_{D4}} \right)^{3/2}\left [ - dt^2+  \delta_{ij}dx^{i}dx^j \right ]
+\left( \frac{u}{R_{D4}} \right)^{3/2} \left [f(u) + \left( \frac{R_{D4}}{u} \right)^{3} \frac{{u'}^2}{f(u)}\right ]dx_4^2 \nonumber \\ \cr
 &+&
 \left( \frac{R_{D4}}{u} \right)^{3/2} u^2 d\Omega_4^2
\eea
where $u'=du/dx_4$.
  Substituting the determinant of the induced metric
and the dilaton into the DBI action, we obtain
(ignoring the factor of $N_f$ which multiplies all the D8-brane
actions that we will write) :
\bea
\label{eightaction}
S_{DBI} =  T_8 \int dt d^3 x d x_4 d^4\Omega e^{-\phi}
\sqrt{-\det(\hat g)}
        =  \frac{\hat T_8 }{g_s}\int  dx_4 u^4 \sqrt{f(u)+
\left( \frac{R_{D4}}{u} \right)^{3} \frac{{u'}^2}{f(u)}},
\eea
where $\hat g$ is the induced metric (\ref{induced}) and $\hat T_8$ includes the
 outcome of the integration over all the coordinates apart from $d
 x_4$.
 The simplest way to solve the equation of motion is by using the
 Hamiltonian of the action (\ref{eightaction}), which is conserved
 (independent of $x_4$) :
\be\label{D8profile} \frac{u^4 f(u)}{\sqrt{f(u)+ \left(
\frac{R_{D4}}{u} \right)^{3} \frac{{u'}^2}{f(u)}}}= {\rm constant} =
u_0^4 \sqrt{f(u_0)}, \ee
where on the right-hand side of the equation we assumed
that there is a point $u_0$ where the profile $u(x_4)$ of the brane has a
minimum, $u'(u=u_0)=0$\footnote{This type of analysis was done
  previously for
  Wilson line configurations. See, for instance,
  \cite{BISY2}.}.
We need to find the solution in which
as $u$ goes to infinity, $x_4$ goes to the values $x_4=0,L$; this
implies
\be
\int dx_4 = 2 \int {\frac{du}{u'}} = L
\label{forl}
\ee
with $u'$ given (as a function of $u$) by
(\ref{D8profile}) (note that $u$ is a double-valued function of $x_4$
in these configurations, leading to the factor of two in (\ref{forl})).
The form of this profile of the D8 brane is drawn in figure
\ref{D8-barD8SS}(a). Plugging in the value of $u'$ from (\ref{D8profile})
we find
\bea
\label{lzerotemp}
L&=&\int dx_4 = 2\int_{u_0}^{\infty} \frac{du}{u'}=
2 R^{3/2}_{D4}
\int_{u_0}^\infty du \frac{1}{f(u)u^{3/2}\sqrt{ \frac{f(u) u^8}{f(u_0) u_0^8}
-1}} \nonumber \\ \cr
&=&  \frac{2}{3} \left( \frac{R^3_{D4}}{u_0} \right)^{1/2}
\sqrt{1-y_K^3}\int_0^1 dz \frac{z^{1/2}}{(1-y_K^3 z)
\sqrt{1-y_K^3 z -(1-y_K^3)z^{8/3}}},
\eea
where $y_K \equiv u_{K} / u_0$.
Small values of $L$ correspond to large values of $u_0$. In this limit
we have $y_{K}\ll 1$ leading to $L\propto \sqrt{R_{D4}^3/u_0}$.
For general values of $L$ the dependence of $u_0$ on $L$ is more complicated.

%%%%%%%%%%%%%%%%%%%%
There is a simple special case of the above solutions, which occurs
when $L = \pi R_x$, namely the D8-branes and anti-D8-branes lie at
antipodal points of the circle. In this case the solution for the
branes is simply $x_4(u) = 0$ and $x_4(u)=L=\pi R_x$, with the two branches
meeting smoothly at the minimal value $u=u_0=U_{K}$ to join
the D8-branes and the anti-D8-branes together.
%It can be seen that for $u_0=u_\Lambda$, $x_4$ approaches
%asymptotically the value $x_4(\infty) =\frac{\delta x_4}{4}$. In fact
%$x_4(u) = \frac{\delta x_4}{4}$ is a solution of the equation of
%motion.
This type of antipodal solution is drawn in figure
\ref{D8-barD8SS}(b).  It was shown in \cite{Sakai:2004cn} that this classical
configuration is stable, by analyzing small fluctuations around this
configuration and checking that the energy density associated with
them is non-negative.

In  general for $L<\pi R_x$, there is a family of
 smooth configurations characterized   by $L$ or by the minimal value of $u$, $u_0$.
This class of solution is shown in \ref{D8-barD8SS}(a)

The Sakai-Sugimoto model has 3 dimensionful parameters : $\lambda_5$, $L$
and $R_x$, and gravity is reliable whenever $\lambda_5 \gg R_x$. The physics
depends on the two dimensionless ratios of these two parameters;  In the gravity limit
the mass of the (low-spin) mesons is related to $1/L$\cite{Aharony:2006da} while the mass of
the (low-spin) glueballs is related to $1/R_x$. As discussed above, in the limit
$\lambda_5 \ll R_x$ this theory approaches (large $N_c$) QCD at low energies.
This remains true also after adding the flavors, at least when $L$ is of
order $R_x$.

The thermal phases of the model where analyzed in \cite{Aharony:2006da,Peeters:2006iu}. The back-reaction
of the flavor brane at non zero temperature is not addressed in the present paper and will be described in a future publication.
%%%%%%%%%%%%%%%%%%%%%%%%%%%%%%%%%%%%%%%%%%
\subsection{Massive type IIA and 8 branes}
%%%%%%%%%%%%%%%%%%%%%%%%%%%%%%%%%%%%%%%%%%

One expects $p+1$ dimensional objects to naturally couple to a $p+1$ form potential.  Therefore, one expects a D8 brane to couple to a nine form potential.  Conventional type IIA supergravity has no such form, and so some modification of the theory is necessary to describe the backreaction of D8 branes.  This extension was first found by Romans \cite{Romans:1985tz}, and then further generalized to admit localized D8 solutions in \cite{Bergshoeff:1995vh}.  The relevant kappa symmetric worldvolume actions were constructed in \cite{Bergshoeff:1996tu}.  Further studies of D8 (-D$p$) brane backgrounds (systems) are discussed in \cite{Massar:1999sb,Janssen:1999sa,Brandhuber:1999np,Mihailescu:2000dn,Witten:2000mf,Bergshoeff:2001pv}.

There exists another massive type IIA, constructed in \cite{Howe:1997qt}.  This and the Romans' type
IIA were shown to be the only ``Higgs type'' supersymmetric extensions of massless type IIA in \cite{Tsimpis:2005vu}, although a third was suggested in \cite{Howe:1997qt}.  The massive type IIA given in \cite{Howe:1997qt} does not admit localized supersymmetric eight-branes, and so we restrict our attention to
the theory of Romans \cite{Romans:1985tz,Bergshoeff:1995vh} which we simply refer to as massive type IIA.

The bosonic part of the action of the massive type IIA supergravity takes the form\footnote{
We use the notation of chapter 12 of Polchinski
\cite{Polchinski:1998rr}}
\bea
S_{IIA_M}&=&S_{NS}+S_R+S_{CS}+S_M \nn \\
S_{NS}&=&\frac{1}{2 \kappa_{10}^2}\int d^{10}x \sqrt{-g}e^{-2\phi}\left(R+4 \partial_\mu \phi \partial^\mu \phi -\frac{1}{2\cdot 3!}H_3\cdot H_3\right) \nn\\
S_R &=& -\frac{1}{2 \kappa_{10}^2}\int d^{10}x \sqrt{-g}\left(\frac{1}{2\cdot 2!}\tilde F_2 \cdot \tilde F_2 + \frac{1}{2 \cdot4!}\tilde{F}_4 \cdot \tilde{F}_4\right) \label{MIIAaction}\\
S_{CS}&=& -\frac{1}{2\kappa_{10}^2}\int d^{10}x \fft12\; \frac{1}{2!\cdot 4!\cdot 4!}\epsilon^{\mu_1\cdots \mu_{10}}B_{\mu_1 \mu_2} \hat F_{\mu_3 \cdots  \mu_6} \hat F_{\mu_7 \cdots \mu_10}\nn \\
S_M&=& -\frac{1}{2\kappa_{10}^2}\int d^{10}x \sqrt{-g} \fft12 M^2+\frac{1}{2\kappa_{10}^2}\int M F_{10} \nn
\eea
where $\cdot$ denotes contraction of indices with inverse metrics, and $\epsilon$ is antisymmetric in indices and takes values $\pm 1$.  In the above action $F_{n+1}=d A_{n}$ and
\bea
\tilde{F}_2&&=F_2+MB_2 \nn \\
\hat{F}_4&&=F_4+\fft12M B_2\wedge B_2  \\
\tilde{F}_4&&=F_4-A_1\wedge H_3+\fft12M B_2\wedge B_2. \nn
\eea
One notes that
from the above definitions, $F_2$ may be absorbed completely by a shift in $B_2$, but only when $M\neq 0$.  One
views this as a ``Higgsing'' where the degrees of freedom associated with $F_2$ become the longitudinal modes
of a massive $B_2$.  The equation of motion for $F_2$, therefore, must only be imposed in the massless limit.  In
 appendix A, we include the equation of motion associated with $A_1$ so that an $M\rightarrow 0$ limit is clear.  In the next section we turn to including sources in the action.

%%%%%%%%%%%%%%%%%%%%%%%%%%%%%%%%%%%%%%%%%
\section{Backreaction of D8 branes}
%%%%%%%%%%%%%%%%%%%%%%%%%%%%%%%%%%%%%%%%%

In this section, we will find the equations of motion that govern the D4-D8 systems of interest, including
the contribution from the DBI + CS brane action.  We present our ansatz, and the perturbative parameter we will use to linearize the equations, and finally present the separated linearized equations.  Further, we find the small coordinate transformations that leave the form of our ansatz unchanged (to the order we are working): these are gauge transformations of the linearized equations.

\subsection{Finding the equations: ansatz and separation}

For the remainder of the paper, we will be concerned with D4-D8 systems, and because neither of these
branes source (directly) either $A_\mu$ or $B_{\mu \nu}$, we set them to zero.
After truncation to the $A_{\mu}=0,B_{\mu \nu}=0$ sector, the equations of motion
for the type IIA massive supergravity are the following:
\bea\label{EOMmassiveIIA}
R_{\mu \nu}+2\nabla_{\mu}\nabla_{\nu}\phi &=& \frac{e^{2 \phi}}{2!4!}
\left(4F_{\mu}\,^{\alpha_2\alpha_3\alpha_4}F_{\nu \alpha_2\alpha_3\alpha_4} - \fft12 g_{\mu \nu} \left(F_{(4)}\right)^2\right)
-\frac{e^{2 \phi}}{4}g_{\mu \nu}M^2 \nn \\
g^{\mu \nu}\nabla_{\mu}\nabla_{\nu}\phi&=&g^{\mu \nu}\left(\nabla_{\mu}\phi \right)\left(\nabla_{\nu}\phi\right) -\fft14 R \nn \\
\nabla_{\alpha_1}F^{\alpha_1 \alpha_2 \alpha_3\alpha_4} &=& 0  \\
\partial_{\mu} M &=&0 \nn \\
M &=& *F_{(10)} \nn \\
0 &=&\frac{1}{\sqrt{-g}}\epsilon^{\beta_1 \beta_2 \alpha_1 \alpha_2 \alpha_3 \alpha_4 \rho_1 \rho_2 \rho_3 \rho_4}
F_{\alpha_1 \alpha_2 \alpha_3\alpha_4}F_{ \rho_1 \rho_2 \rho_3 \rho_4}. \nn
\eea
Note that the equation of motion for $A_1$ in the appendix is trivially satisfied.  Again,
one must only impose it's equation
of motion in the massless limit.
However, the equation of motion for $B_{\mu \nu}$ imposes a constraint, arising from the
Chern Simons term $B\wedge F_4 \wedge F_4$, on the four form (the last of the above equations).
This constraint is easily satisfied for simple 4-form field
strengths.  Also, note that we have used the dilaton equation of motion (EOM) to eliminate $R$ from the Einstein equation.
This will be important below when we derive the equations when a brane source is present.

We now turn to the modification of the  equations of motion (\ref{EOMmassiveIIA}) by adding
\be
-K_{p}\left(\int d^{p+1}\xi e^{-\phi} \sqrt{-g_p} + \int A_{p+1} \right)
\ee
to the action.
Here we use $g_p$ to denote the pullback metric on the $p+1$ dimensional
submanifold defined by $X^\alpha(\xi)$,
\be
\left(g_{p}\right)_{a b}= \frac{\partial X^{\mu}}{\partial \xi^{a}}\frac{\partial X^{\nu}}{\partial \xi^{b}}g_{\mu \nu}
\ee
and $K$ is the appropriate constant involving
the p-brane tension.  In this action, we assume that it is consistent to set the
world volume $U(1)$ gauge field to zero, as well as ignoring any additional Chern Simons terms
(which  is appropriate for the cases we wish to consider).

There are
two types of fields in this action: those that represent open string degrees of freedom
(e.g. $X^{\alpha}(\xi)$); and those representing closed string degrees of freedom
(e.g. $g_{\alpha \beta}$).  Of course when varying with respect to the closed string
degrees of freedom, 10D delta functions appear, of which $p+1$ are integrated leaving
behind a $(10-(p+1))$ dimensional delta function source term, as we should expect.   Varying the
above action with respect to the $p+1$ form potential adds a delta function source
to the form fields equation of motion of the generic form
\be
\frac{1}{(2\kappa_{10}^2 )}(d*F)-K_{p}\delta^{10-p-1}\frac{(\epsilon_{10}\cdot\epsilon_{p+1})}{(p+1)!}=0.
\ee
Note that the product $\epsilon_{10}\cdot\epsilon_{p+1}$ is sensitive to the orientation of the
submanifold defined by $X^{\alpha}$.  For example, in the case of the SS model with the antipodal embedding
of the D8 branes, there is both a positive delta function and a negative delta function in $x_4$ accounting
for the
orientation reversal of the brane (it is oriented in the $\pm u$ direction).
For D8 branes, however, one should replace $d*F$ with $dM$ because $M$
is the term that appears with $F_{10}$ in the action.
Hence, $M$ is in fact piecewise constant in backgrounds with localized D8, as we will see below.

We restrict ourselves to embedding functions of the form $X^{a}(\xi)=\xi^{a}$
and the remaining $X^{i}$ are arbitrary constants.
Varying the full action with respect to the dilaton and graviton is now straightforward,
and the equations of motion are
\bea
R_{\mu \nu}+2\nabla_{\mu}\nabla_{\nu}\phi - \frac{e^{2 \phi}}{2!4!}
\left(4F_{\mu}\;^{\alpha_2\alpha_3 \alpha_4}F_{\nu \alpha_2 \alpha_3 \alpha_4} - \fft12 g_{\mu \nu} \left(F_{(4)}\right)^2\right)
+\frac{e^{2 \phi}}{4}g_{\mu \nu}M^2 &&\nn \\
+\frac{K_82\kappa_{10}^2}{2} e^{\phi} \frac{\sqrt{-g_p}|_{\xi^i=x^i}}{\sqrt{-g}}
\left(g_p^{a b} \frac{\partial X^{\alpha}}{\partial \xi^a}\frac{\partial X^{\beta}}{\partial \xi^b}|_{\xi^i=x^i}\;g_{\alpha \mu} g_{\beta \nu} - \fft12 g_{\mu \nu}\right)\delta\left(x^{\alpha}-X^{\alpha}(x)\right) &=&0 \nn \\
R+4g^{\mu \nu}\nabla_{\mu}\nabla_{\nu}\phi-4g^{\mu \nu}\left(\nabla_{\mu}\phi \right)\left(\nabla_{\nu}\phi\right)
-\frac{K_82\kappa_{10}^2}{2} e^{\phi} \frac{\sqrt{-g_p}|_{\xi^i=x^i}}{\sqrt{-g}}
\delta\left(x^{\alpha}-X^{\alpha}(x)\right) &=&0 \nn \\
\frac{1}{(2\kappa_{10}^2)}d M-K_8\delta\left(x^{\alpha}-X^{\alpha}(x)\right)\frac{(\epsilon_{10}\cdot\epsilon_{9})}{9!}&=&0 \nn \\
&& \kern-8em *F_{(10)} = M.
\eea
with other equations of motion left unchanged.  The delta functions appearing above
may be simplified by taking them to be functions only of $x_4$, $\delta\left(x^a-X^a(x)\right)\equiv\Delta(x_4)$,
which is appropriate for the antipodal embedding in the Sakai Sugimoto model.
As expected, only the RR couplings to the
branes are sensitive to the orientation of the branes.  Further, the epsilons appearing above take
values $\pm 1$, and do not contain factors of $\sqrt{-g}$.
\footnote{As a simple check of the above signs and numerical factors, one can simply check the following.  The coefficient
in front of $g_p^{ab}\partial_a X^{\alpha} \partial _b X^{\beta} g_{\alpha \mu} g_{\beta \nu}$ can be checked against that of $R_{\mu \nu}$.
Before using the dilaton equation of motion, the Einstein equations contain $-\fft12 R g_{\mu \nu}$, and this coefficient
must match that of $g_p^{ab}\partial_a X^{\alpha} \partial _b X^{\beta} g_{\alpha \mu} g_{\beta \nu}$, except that one multiplies the
latter by an additional $-K_8 2 \kappa_{10}^2 \sqrt{-g_p}/\sqrt{-g}$.  This is because they are obtained from similar terms
in the action, $\frac{1}{2\kappa_{10}^2}\sqrt{-g}e^{-2 \phi}R$ and $-K_8e^{-\phi}\sqrt{-g_p}$ respectively.  The factor in the dilaton equation is also obtained similarly, as the coefficient of R is obtained by varying $\frac{1}{2\kappa_{10}^2}\sqrt{-g}e^{-2 \phi}R$ w.r.t. $\phi$ and the delta function coefficient is obtained by varying $-K_8 \sqrt{-g}e^{- \phi}$, and so a factor of $1/2$ arises.  The second part of the delta function term
in Einstein's equations is similarly found by checking that one adds $\fft12 g_{\mu \nu}$ times that of the dilaton term.}

The tensor structure of the Einstein equations can be easily read:  the delta function strength is
proportional to the metric and dilaton, and comes with a $+$ sign for a direction
along the D8 brane, and comes with a $-$ sign for those not along the D8 brane.  Although
above we have written the effect of a D8 brane, the above arguments work also
for an arbitrary $p$ brane: it simply changes
which RR form field equation of motion gets the delta function source, and how many directions
of the Einstein's equations get $(-)$ vs. $(+)$ delta functions.

For the remainder of this work, we will take the solution to the $M$ and $F_{10}$ equations of motion
to be
\bea
M&=&\pm 2\kappa_{10}^2 K_8/2=\pm \frac{N_f}{4\pi \ell_s}\equiv \pm\frac{Q_f}{g_s} \nn \\
*F_{10}&=&M
\eea
where the $+$ is used on one side of the D8, and $-$ is used on the other. \footnote{This terminology
only makes sense for branes of dimension $D-1$, as such branes split the space into disjoint regions.}

The remainder of the paper will be devoted to solving the remaining equations of motion perturbatively,
and the perturbative control parameter will be explained shortly.  However, at this point
an important note is in order: delta functions in
codimension $10-(p+1)\neq 1$ have singularities at the location of the delta function.  Codimension
one is special in that the Green's function is of the form $|x|$ \footnote{we refer to this behavior as a cusp},
and hence is finite at the
source.  We therefore expect that the perturbative approach is most natural for D8 branes, as the
back reaction can be made small, even close to the brane.  Hence, the perturbative approach that we take
may not be as natural for higher codimension branes.

To define the small parameter in our expansion, we make the following observations.  Given the solution of the $M,F_{10}$ sector,
the new (relative to the massless IIA equations) terms in the equations of motion come with the
powers of $2\kappa_{10}^2 K_8 e^{\phi}\sim g_s N_f$.  This is what we shall use
as a control parameter for our perturbative expansion.  From
now on, we will simply take $2\kappa_{10}^2 K_8/2 \equiv Q_f/g_s$ as the definition
of our small parameter $Q_f$.  Another way of phrasing this is that in the holographic
limit, there is a scale $R_c$ such that $g_s\sim (R_c /l_s)/N_c$ where $R_c/ l_s\gg 1$ is large but held fixed.  Hence, one may view our perturbation
as limit on $g_s N_f\sim N_f/N_c$, which is the basis for the probe approximation.

To solve the equations, we will still need to take an ansatz, and we motivate it as follows.
The eight brane doesn't directly couple to $F_4$, and further the $SO(5)$ symmetry of the
4 sphere is not broken for this brane configuration.  Hence, we take that there is no
change in $F_4$ to leading order in $Q_f$.  Further, because we take the solution
$M\propto \pm Q_f/g_s$, so that in the Einstein equations, the term $g_{\mu \nu} M^2 e^{2\phi}\sim \mathcal{O}\left(Q_f^2\right)$
may be ignored.

Therefore, we assume that only metric and dilaton perturbations are necessary, so we take a general ansatz of the form
\bea
ds^2&=&e^{2A(u,x_4)}\left(-dt^2+dx_i^2\right)+e^{2B(u,x_4)}dx_4^2+e^{2G(u,x_4)}du^2+e^{2C(u,x_4)}d\Omega_4^2 \nn \\
\phi(u,x_4)&=&\frac{1}{2}\hat{\phi}(u,x_4)+2A(u,x_4)+2C(u,x_4) \label{deff}\\
F_{(4)}&=&Q_c \Omega_4 \nn
\eea
where $Q_c=3 R_4^3/g_s=3 \pi N_c l_s^3$, and $\Omega_4$ is the volume form of the unit four sphere.  It is clear that the $F_4$ equations
are trivially satisfied: $dF_4=0$ because $\Omega_4$ is closed, and $d*F_4=0$ because $*F_4$ is some function of $x_4$ and $u$ times $du\wedge dx_4 \wedge_i dx_i$, and is therefore closed.

In the above, we will expand the above functions as
\bea
&&A(u,x_4)=A_0(u)+Q_f A_1(u,x_4) \quad B(u,x_4)=B_0(u)+Q_f B_1(u,x_4) \nn\\
&&C(u,x_4)=C_0(u)+Q_f C_1(u,x_4) \quad G(u,x4)=G_0(u)+Q_f G_1(u,x_4) \\
&&\phantom{C(u,x_4)=C_0(u)+Q_f} \ph(u,x_4)=\ph_0(u)+Q_f \ph_1(u,x_4)\nn
\eea
where the $0$ subscripted functions are solutions of the $Q_f=0$ equations.  We linearize and explain how to
separate them for the Sakai Sugimoto model in appendix B, and summarize the results here.
One must solve the decoupled system
\bea
&&3\pau^2F_1+\frac{3(4u^3-U_K^3)\pau F_1}{u(u^3-U_K^3)}
+\frac{g_s Q_c u^3 \pax^2 F_1}{(u^3-U_K^3)^2} \nn \\
&& \quad \quad  -\frac{54 u F_1}{(u^3-U_K^3)}+\frac{2u(Q_c g_s)^{\fft23}\sqrt{3}}{(u^3-U_K^3)}\sqrt{\frac{u}{(Q_c g_s)^{\fft13}\left(1-\frac{U_K^3}{u^3}\right)}}\Delta=0 \label{def1}\\
\phantom{x} &&  \nn \\
\phantom{x} &&  \nn \\
&& 3\pau^2F_2+\frac{3(4u^3-U_K^3)\pau F_2}{u(u^3-U_K^3)}
+\frac{ g_s Q_c u^3 \pax^2 F_2}{(u^3-U_K^3)^2} \nn \\
&& \quad \quad -\frac{4u(Q_c g_s)^{\fft23}\sqrt{3}}{(u^3-U_K^3)}\sqrt{\frac{u}{(Q_c g_s)^{\fft13}\left(1-\frac{U_K^3}{u^3}\right)}}\Delta=0 \label{def2}\\
\phantom{x} && \nn \\
\phantom{x} && \nn \\
&&-4 \pau^2 \ph_1 -\fft43\frac{u^3Q_cg_s \pax^2 \ph_1}{\left(u^3-U_K^3\right)^2}-\frac{2\left(u^3-7U_K^3\right)\pau \ph_1}{u\left(u^3-U_K^3\right)}-\frac{36 u U_K^3 \ph_1}{\left(u^3-U_K^3\right)^2} \nn \\
&& \quad \quad+\frac{\fft43 u(Q_c g_s)^{\fft23}\sqrt{3}}{(u^3-U_K^3)}\sqrt{\frac{u}{(Q_c g_s)^{\fft13}\left(1-\frac{U_K^3}{u^3}\right)}}\Delta=0 \label{deph}
\eea
where
\be
\Delta=
\begin{cases}
\delta\left(x_4 \right)+\delta\left(x_4-\pi R_x \right) & \mbox{if $x_4= x_4+2\pi R_x$} \\
\delta\left(x_4\right) & \mbox{if $x_4$ non-compact}
\end{cases}
\ee
and then identify the physical degrees of freedom
\bea
A_1 &=& -\fft15 F_1 + \frac{1}{10} F_2 -\frac{3}{10} \ph_1 \nn \\
C_1 &=& \fft{1}{10} F_1 + \fft15 F_2 - \fft{1}{10} \ph_1 \nn \\
B_1=G_1&=&-\fft15F_1(u,x_4)+\fft15 F_2(u,x_4)-\fft{1}{10} \ph_1(u,x_4)\nn \\
&& \quad -\fft25 u \pau \ph_1(u,x_4)+\frac{3}{5}\frac{\ph_1(u,x_4) U_K^3}{(u^3-U_K^3)} \label{findBG}\\
\phi_1&=&\fft12 \ph_1 +2A_1+2C_1. \nn
\eea
Above, we have made the obvious notation that $\phi_1$ is the first order correction to the {\it physical}
dilaton.  One may read off the combined solution by plugging in these to the equations (\ref{fexp}) in the appendix.

%%%%%%%%%%%%%%%%%%%%%%%%%%%%%%%%%%%%%%%%%%
\subsection{Gauge Freedom}
%%%%%%%%%%%%%%%%%%%%%%%%%%%%%%%%%%%%%%%%%%

Here we identify the gauge (coordinate transformation) freedom as those transformations in $u$ and $x_4$ that leave the metric diagonal (preserves the form of our ansatz).  Indeed,
\bea
\ph_1\rightarrow && \ph_1 - \fft52 \frac{\lambda(u,x_4)}{u} \nn \\
A_1\rightarrow && A_1 + \fft34 \frac{\lambda(u,x_4)}{u} \nn \\
B_1\rightarrow && B_1 + \fft34 \frac{\lambda(u,x_4)}{u} +\fft32 \frac{U_K^3\lambda(u,x_4)}{u(u^3-U_K^3)}\nn \\
 &&\quad \quad \quad -\fft13 \int{\frac{Q_c g_s (\pax \pax \lambda(u,x_4))}{u^3\left(1-\frac{U_K^3}{u^3}\right)^2}du} \label{gague} \\
G_1\rightarrow && G_1 - \fft34 \frac{\lambda(u,x_4)}{u}-\fft32 \frac{U_K^3\lambda(u,x_4)}{u(u^3-U_K^3)}+\pau \lambda(u,x_4) \nn \\
C_1\rightarrow && C_1 + \fft14 \frac{\lambda(u,x_4)}{u} \nn
\eea
leaves all equations of motion unchanged, and is exactly a coordinate change in $u$ and $x_4$, namely
\bea
u&&\rightarrow u+\lambda(u,x_4) \\
x_4&&\rightarrow x_4-\fft13 \int{\frac{Q_c g_s (\pax \lambda(u,x_4))}{u^3\left(1-\frac{U_K^3}{u^3}\right)^2}du} \nn
\eea
Hence, one of the degrees of freedom above is pure gauge.  However, there is
an added complication.  If we
eliminates $\ph_1$ using such a gauge transformation, the cusp in $\ph_1$ generates a delta function in the
gauge transformation for $B_1$, and hence $B_1$ is no longer a smooth function: it contains a delta function.  Hence,
we conclude that for the unsourced equations one may choose which degree of freedom to eliminate, but in the sourced
equations, only $B_1$ may be eliminated.  However, as shown in appendix B, it is more convenient to not eliminate $B_1$ completely, but rather to take $B_1=G_1$ as the choice.

Given the above transformations, we can immediately see that $F_1$ and $F_2$ of the last subsection are gauge independent.
The remaining gauge dependent quantities $\ph_1$, $B_1$ and $G_1$ do not admit a gauge independent combination.
Further, given the equations (\ref{findBG}), only the equation $B_1=G_1$ is not gauge covariant.  Therefore,
we will sometimes refer to this as a gauge fixing.

%%%%%%%%%%%%%%%%%%%%%%%%%%%%%%%%%%%%%%%%
\section{Solutions: the linearized backreaction}
%%%%%%%%%%%%%%%%%%%%%%%%%%%%%%%%%%%%%%%%%

Here we will analyze the differential equations of the last section in three separate cases:
\begin{enumerate}
\item $U_K=0$ decompactification limit:  In this case we take $U_K=0$, in a limiting sense of the background.  In this limit $R_x^2 \propto R_{(D4)}^3/U_K$ becomes infinite, and so $x_4$ decompactifies.
\item $U_K=0$, $x_4$ compactified: In this case, we note that while the $U_K=0$ limit has decompactified $x_4$, one still has the isometry $x_4\rightarrow x_4+$ constant.  Hence, one may orbifold by this isometry and compactify $x_4$.  We will parameterize this compactification using the same radius, $R_x^2=(4/9)R_{(D4)}^3/U_K$.  One way to think of this parametrization is that we have taken the spacetime to be that of
    $U_K=0$ while still requiring that $x_4$ is compactified: we choose to parameterize the compactification of $x_4$ such that we may compare easily to the $U_K\neq 0$ case.  In this way, we have taken the spacetime to be the cylinder to which the cigar asymptotes, and so this analysis gives the $u\gg U_K$ behavior of the $U_K \neq 0$ case.  In all compact $x_4$ cases, we will be considering the antipodal embedding, $L=\pi R_x$, which for concreteness we parameterize by the embedding $x_4=0,\pi R_x$.
\item $U_K\neq 0$:  In this case, we analyze the equations as is.  We make some basic comments about the nature of the Fourier transformed equations, and note that the point $u=U_K$ is a regular singular point, and hence a finite convergent series about this point exists.  We expand the solution about the tip of the cigar.
\end{enumerate}

%%%%%%%%%%%%%%%%%%%%%%%%%%%%%%%%%%%%%%%%%%
\subsection{$U_K=0$ decompactification limit}
%%%%%%%%%%%%%%%%%%%%%%%%%%%%%%%%%%%%%%%%%%

In the $U_K\rightarrow 0$ limit, the differential equations become
\bea
3 \pau^2 F_1 +\frac{12}{u}\pau F_1-\frac{54}{u^2}F_1 + \frac{Q_c g_s}{u^3} \pax^2 F_1+
\frac{2 (3Q_c g_s)^{\fft12}\Delta}{u^{\fft32}}&=&0 \nn \\
3 \pau^2 F_2 +\frac{12}{u}\pau F_2 + \frac{Q_c g_s}{u^3} \pax^2 F_2-
\frac{4 (3Q_c g_s)^{\fft12}\Delta}{u^{\fft32}}&=&0 \\
-4 \pau^2 \ph_1 - \frac{2}{u}\pau \ph_1 -\frac{4 Q_c g_s}{3 u^3} \pax^2 \ph_1
+\frac{4 (3Q_c g_s)^{\fft12}\Delta}{3u^{\fft32}}&=&0. \nn
\eea
In the above equations, we take all functions to be functions of the form
\bea
F_i(u,x_4)&=&u K_i(q) \quad q=\frac{x_4 \sqrt{u}}{(Q_c g_s)^{\fft12}} \nn \\
\ph_1(u,x4)&=& u K_3(q).
\eea
This has the effect of changing the delta function in $x_4$ into a delta function in $q$ as $\Delta(x_4)=\Delta'(q) \sqrt{u/(Q_c g_s)}$.  Further, we take just a single brane so that $\Delta'=\delta(q)$.
Taking the resulting equations, and multiplying them by $u$, we obtain ODE's with delta function sources
\bea
\left(\fft34 q^2 +1\right)\paq^2 K_1 +\frac{33}{4}q \paq K_1 -42 K_1 +2 \sqrt{3} \delta(q)=0 \nn \\
\left(\fft34 q^2 +1\right)\paq^2 K_2 +\frac{33}{4}q \paq K_2 +12 K_2 -4 \sqrt{3} \delta(q)=0 \\
\left(\fft34 q^2 +1\right)\paq^2 K_3 +3q \paq K_3 +\fft32 K_3 - \sqrt{3} \delta(q)=0 \nn
\eea
One constructs the delta function solution from the vacuum solution.
The vacuum solutions to these equations may be written
\bea
K_1&=&K_a\left(\frac{1}{42}+\fft12 q^2 +q^4\right) \nn \\
&& \kern-1em + K_b\frac{\left(3402q^{13}+22113q^{11}+57915q^{9}+77220q^{7}+54054q^5+18018q^3+2002q\right)}{(3q^2+4)^{\fft92}} \nn\\
K_2&=&K_c\frac{\frac{280}{3}q\ln\left(\sqrt{3q^2+4}+q\sqrt{3}\right)+\sqrt{3q^2+4}
\sqrt{3}\left(q^6+\frac{19}{3}q^4+\frac{58}{3}q^2-\frac{128}{9}\right)}{(3q^2+4)^{\fft92}} \nn \\
&&+ K_d\frac{\frac{280}{3}q\ln\left(\sqrt{3q^2+4}-q\sqrt{3}\right)-\sqrt{3q^2+4}
\sqrt{3}\left(q^6+\frac{19}{3}q^4+\frac{58}{3}q^2-\frac{128}{9}\right)}{(3q^2+4)^{\fft92}} \nn \\
K_3&=&K_e\frac{q}{3q^2+4}+K_f\frac{1}{3q^2+4}
\eea
To obtain even (in $q\rightarrow -q$) convergent (for large $q$) quantities with cusps, we may construct the combinations
\bea
K_1&=&\frac{-256\sqrt{3}}{1001}\Bigg(-\frac{3402}{3^{\fft92}}\left(\frac{1}{42}+\fft12 q^2 +q^4\right)\nn \\
 &&\kern-2em + \left|\frac{\left(3402q^{13}+22113q^{11}+57915q^{9}+77220q^{7}+54054q^5+18018q^3+2002q\right)}{(3q^2+4)^{\fft92}}\right|\Bigg) \nn\\
K_2&=&2^{10}\sqrt{3}\frac{|q|}{(3q^2+4)^{\fft92}} \label{Ks} \\
&&+N_2\frac{\frac{280}{3}q\ln\left(\frac{\sqrt{3q^2+4}+q\sqrt{3}}{\sqrt{3q^2+4}-q\sqrt{3}}\right)+2\sqrt{3q^2+4}
\sqrt{3}\left(q^6+\frac{19}{3}q^4+\frac{58}{3}q^2-\frac{128}{9}\right)}{(3q^2+4)^{\fft92}} \nn \\
K_3&=&2\sqrt{3}\frac{|q|}{3q^2+4}+N_3\frac{1}{3q^2+4}.\nn
\eea
The above have been written with the cusp solution first, and then an even function that converges (with coefficients $N_i$).  We have not been able to determine physical boundary conditions that fix $N_i$, and so we will leave them arbitrary when possible.

\begin{figure}[ht!]
\centering
\subfloat[$K_1(q)$]{\label{K1q}
    \includegraphics[width=0.3\textwidth,angle=-90]{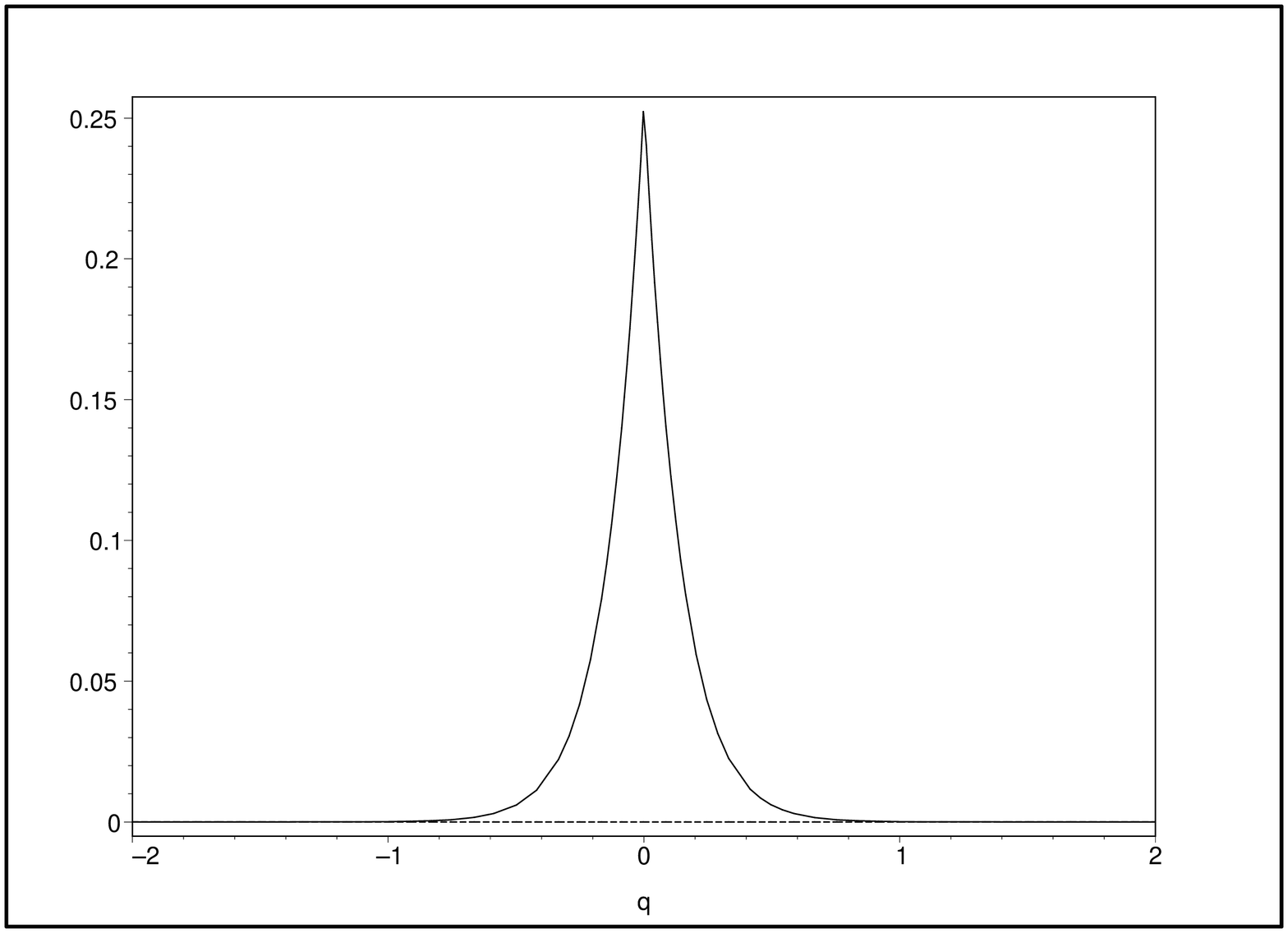}}
\subfloat[$K_2(q)$]{\label{K2q}
    \includegraphics[width=0.3\textwidth,angle=-90]{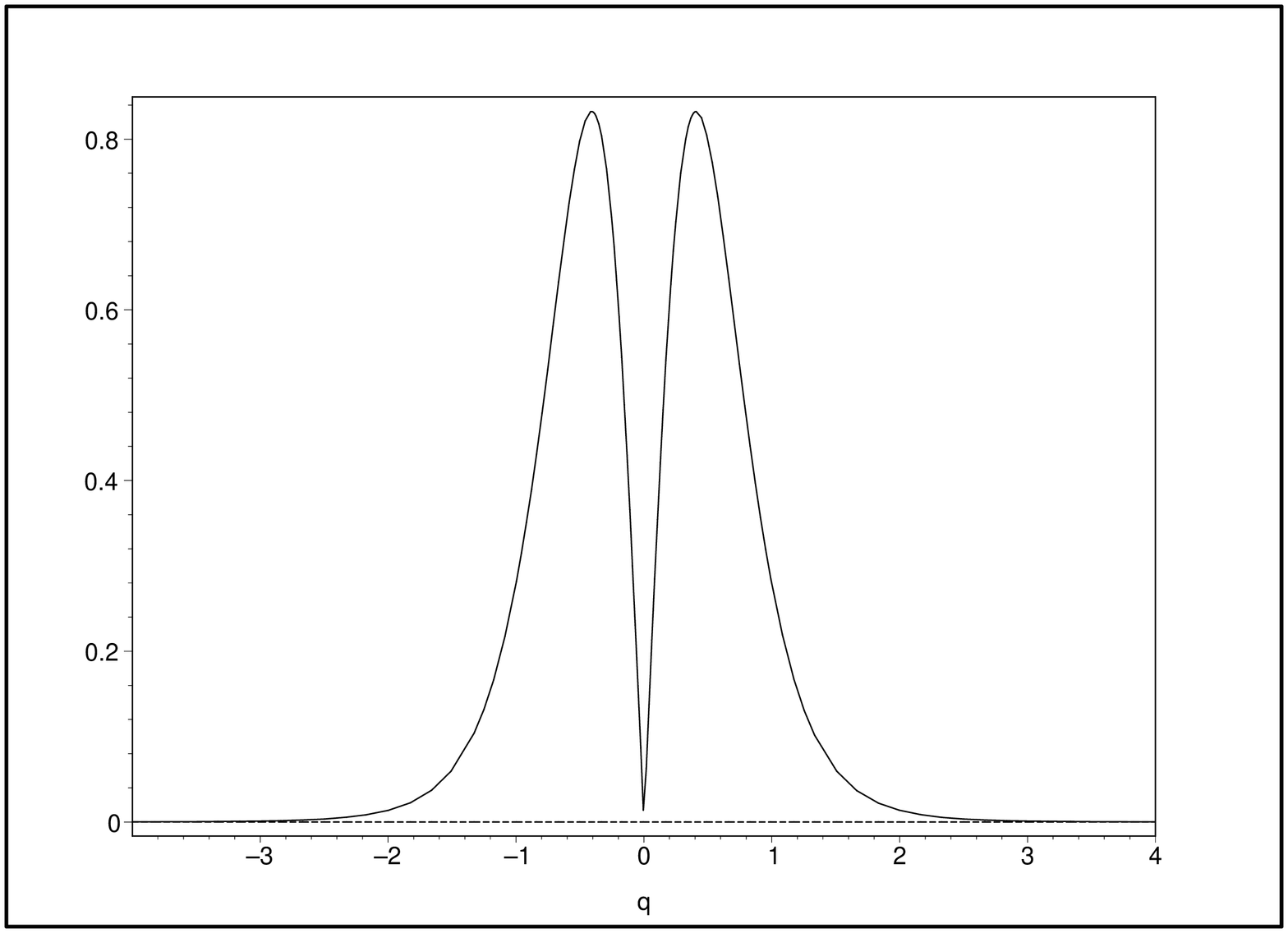}} \\
\subfloat[$K_3(q)$]{\label{K3q}
    \includegraphics[width=0.3\textwidth,angle=-90]{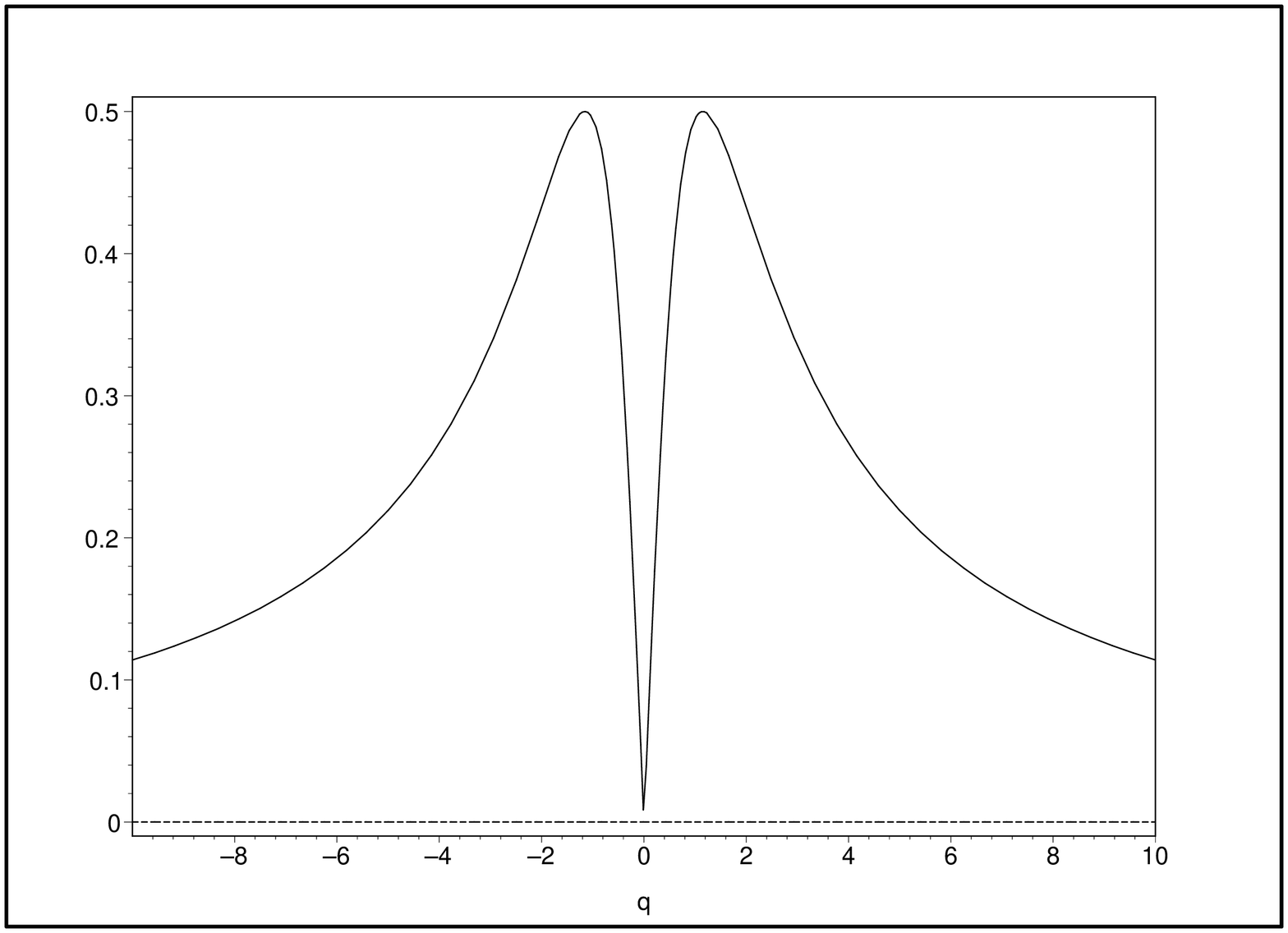}}
\caption{ Graphs of $K_i(q)$, setting $N_2=N_3=0$.}
\label{graphq}
\end{figure}

To graph them, however, we take $N_i=0$ and show these in figure \ref{graphq}.
Further, note that the function $K_3$ has a larger characteristic width, as it only converges as $1/q$. This will be important when we compactify $x_4$.

The height of the above functions grows as $u$ because the peak happens at a fixed value of $q$, giving just a constant contribution times the dressing factor of $u$.  Thus, one expects the perturbative approach to be valid up to $u\ll 1/Q_f$.  This will be generic for later sections as well, as the decompactified behavior emerges at large $u$ in the following sections.

We also wish to characterize the width of the ``spike'' in each
graph.  One way is to make sure that variations happen on scales
larger than string scale.  The slope of the graphs is largest in the
vicinity of the spike, and this slope is determined by its $q=0$
behavior, and is therefore just a constant.  Therefore, the $x_4$
slope is simply $d\phi_i/dx_4=Q_f u^{3/2} R_{D4}^{-3/2}
\times$constant.  The physical length that this corresponds to,
however, is $ds= (u/R_{D4})^{(3/4)}dx_4$, and we require that
$d\phi_i/ds \ll 1/\ell_s$. This gives $u^3\ll
R_{D4}^3/(Q_f^{4}\ell_s^4)\propto R_{D4}^3/(g_s^4 N_f^4)$. Recalling
the conditions above $u\ll 1/Q_f$ and $N_c\gg N_f$, this condition
follows, and so is not a new piece of information.

One may also wish to characterize the width when the linear part is no longer the dominant, and so characterize the width of when other ``features'' become important.  This occurs when the $q$ coordinate becomes order 1, and so translates into $x_4 \propto R_{D4}^{3/2}/u^{1/2}$.  Again, translating this into a physical distance we find $s=R_{D4}^{3/4}u^{1/4}\gg \ell_s$, where we have required that this distance be greater than string scale. This gives a lower bound on $u$, however, it is the same lower bound coming from the Ricci scalar $R\propto 1/(u R^3)^{1/2}\ll 1/\ell_s^2$ for the supergravity approximation.  We see that we trust the supergravity to describe the backreaction above $u\gg \ell_s^4/R_{D4}^3$, and that the perturbative results are good up to $u\ll 1/Q_f$.

Of course one may take the last two constraints on $u$ and turn them into a unitless constraint on the parameters.  We find that this is $1/Q_f\gg \ell_s^4/R_{D4}^3\rightarrow g_s^2 N_f/N_c \ll 1$ which we can see is weaker than the other constraints $N_f/N_c \ll 1$, $g_s \ll 1$.

%To do this, one can simply expand about $x_4=0$ and compare the $Z_1|x_4|$ piece to the $Z_2 x_4^2$ piece (where the %$Z_i$ are functions of u), and find at what value of $x_4$ that $(Z_1|x_4|/(Z_2 x_4^2))= 1$.  One finds that this %happens at a value of $x_4\propto R_{D4}^{3/2}/u^{1/2}$.  This corresponds to a physical length of $ds \propto %R_{D4}^{3/4}u^{1/4}\gg \ell_s$.  This gives a lower bound on $u$, however, it is the same lower bound coming from the %Ricci scalar $R\propto 1/(u R^3)^{1/2}\ll 1/\ell_s^2$ for the supergravity approximation.  We see that we trust the %supergravity to describe the backreaction above a certain $u\gg \ell_s^4/R_{D4}^3$, and that the perturbative results %are good up to a certain value of $u\ll 1/Q_f$.

%%%%%%%%%%%%%%%%%%%%%%%%%%%%%%%%%%%%%%%%%%
\subsection{$U_K=0$ with $x_4$ compactified}
%%%%%%%%%%%%%%%%%%%%%%%%%%%%%%%%%%%%%%%%%%
Here we will examine the $U_K=0$ case with $x_4$ compactified as explained at the beginning of section 4.  However, a few brief words are in order.  We will do this case in two ways: by summing the images from the last section, and by Fourier decomposing them.  These two approaches are complimentary because one expects the sum on images to converge quickly for large $u$ (when the images are well separated), and as we will see, the Fourier modes converge extremely quickly as $u\rightarrow 0$.  Further, we will be considering the antipodal embedding, $L=\pi R_x$, which for concreteness we parameterize by the embedding $x_4=0,\pi R_x$.

\subsubsection*{Sum images of Decompacitification}

Looking at the Fourier transform (below), there is no problem with compactifying
$x_4$, however the solution for $K_3$ above seems to have a problem.  Note that if we take the above $K_3$ and
sum over images in $q$ for some fixed value of $u$, we expect a divergence, as the function only converges
as $1/q$.  However, switching back to the $u,x_4$ language, we find that the $n^{\rm th}$ image of $K_3$
is
\be
\ph_{n,1}=\frac{3\frac{u^{\fft32}}{U_K^{\fft12}}\left|\frac{x_4}{R_x}+\pi n\right|}{\frac{u}{U_K}\left(\frac{x_4}{R_x}+\pi n\right)^2+9}
\rightarrow \frac{3 \sqrt{u}\sqrt{U_K}}{\pi\left|n\right|}
\ee
where the right hand side is it the large $n$ behavior.  This, however, is actually a homogeneous solution to the original differential equation we started with.
This suggests a solution to this difficulty, and we take instead
\be
\ph_{n,1}=\frac{3\frac{u^{\fft32}}{U_K^{\fft12}}\left|\frac{x_4}{R_x}+\pi n\right|}{\frac{u}{U_K}\left(\frac{x_4}{R_x}+\pi n\right)^2+9}
- \frac{3 \sqrt{u}\sqrt{U_K}}{\pi\left|n+C_n\right|}
\ee
where this is just adding a zero mode of the differential equation.
The sum of this function in $n$ converges, as the large $n$ behavior is order $1/n^2$.

Now we make one more final comment.  To fix $C_n$, we require that \newline
$\lim_{u\rightarrow \infty} \ph_1(u,x_4=\pi R_x/2)=0$.  This is roughly requiring that when you are as far away from the branes as possible that the perturbation should be 0.
The solution to this constraint is $C_n=1/2$.  Of course a different set of $C_n$ could be chosen in such a way as to not affect the sum: this, by definition, is unphysical, as none of the field values would change.

\begin{figure}[ht*]
\centering
\subfloat[$F_1(u,x_4)/U_K$]{
    \includegraphics[width=0.3\textwidth,angle=-90]{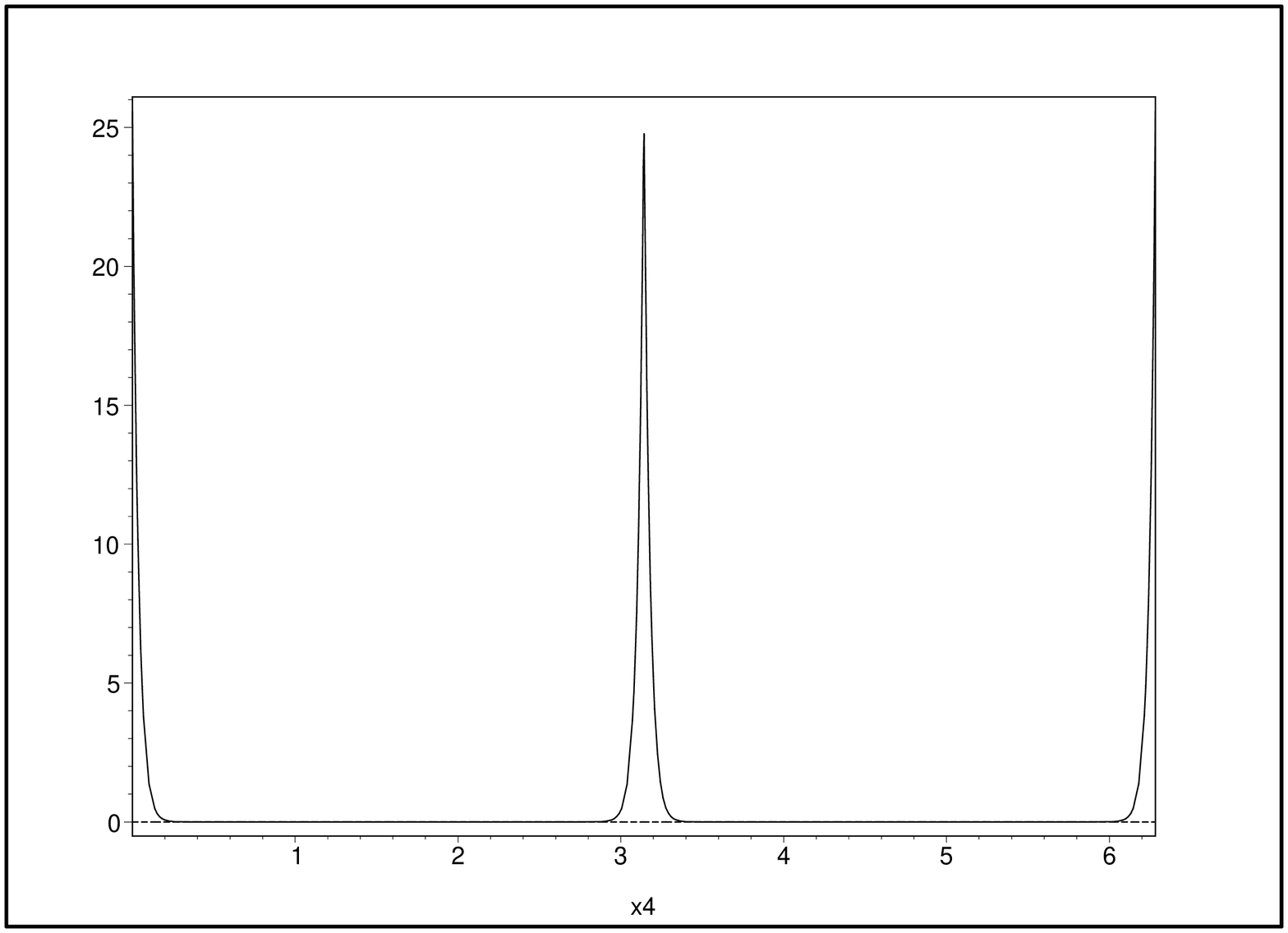}}
\subfloat[$F_2(u,x_4)/U_K$]{
    \includegraphics[width=0.3\textwidth,angle=-90]{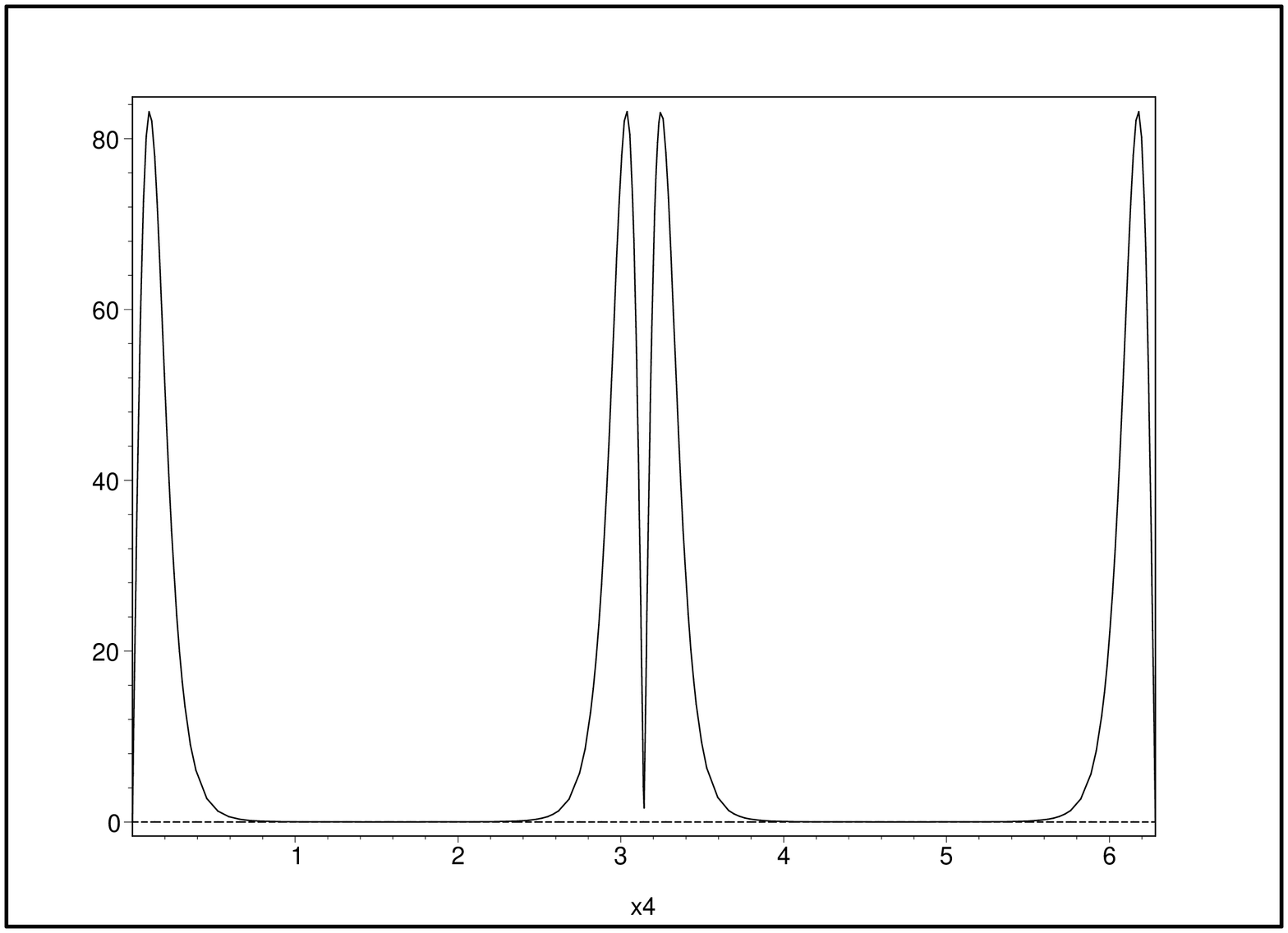}} \\
\subfloat[$\ph_1(u,x_4)/U_K$]{
    \includegraphics[width=0.3\textwidth,angle=-90]{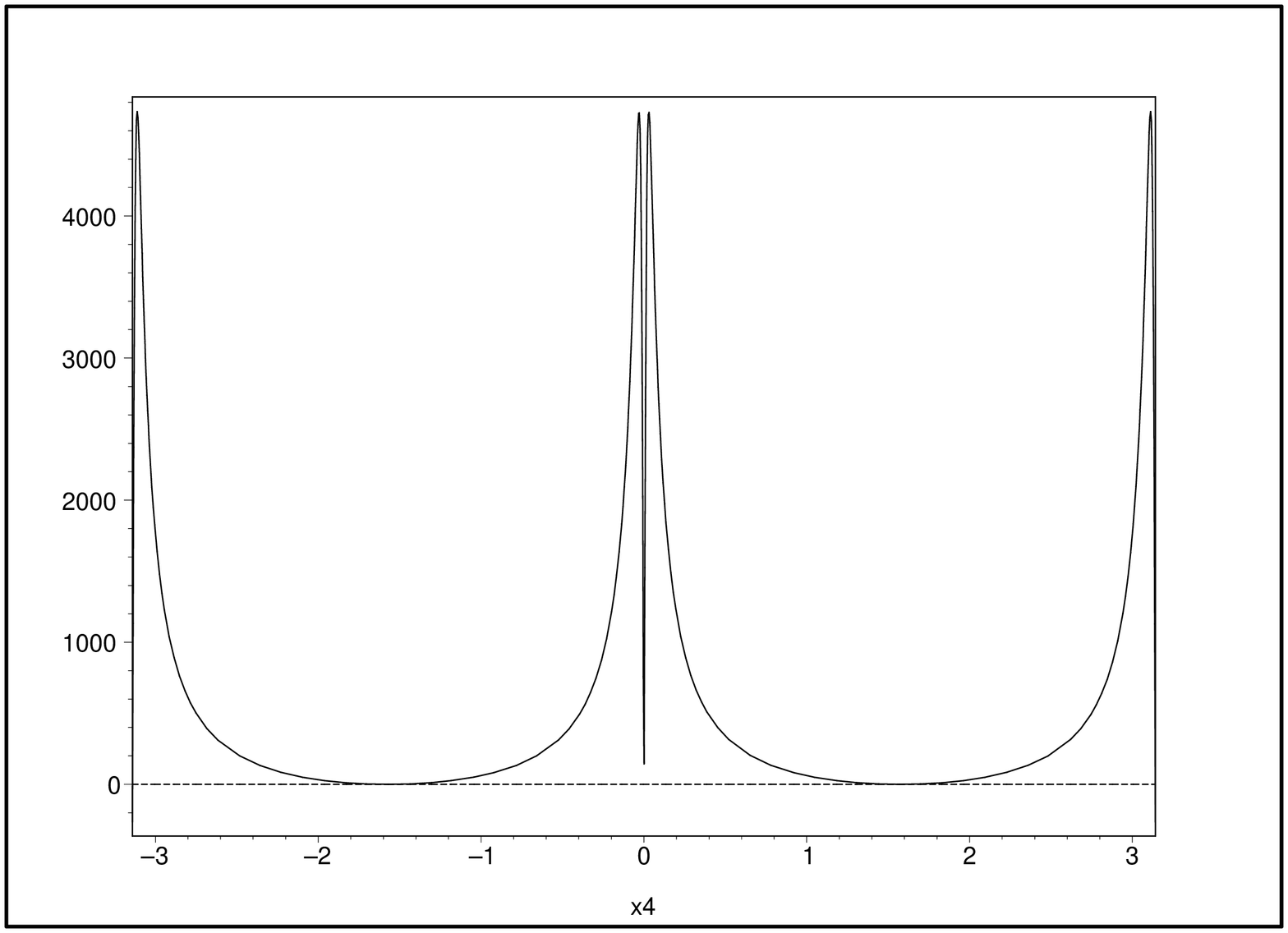}}
\caption{$F_1/U_K$ as a function of $x_4/R_x$ for $u/U_K=100$ using images from $n=-50\cdots 50$,
$F_2/U_K$ as a function of $x_4/R_x$ for $u/U_K=100$, $n=-500\cdots 500$ and
 $\ph_1/U_K$ as a function of $x_4/R_x$ for $u/U_K=10,000$, $n=-500\cdots 500$.}
\label{plotSumImages}
\end{figure}

Therefore, for the compactified case, we take
\bea
\ph_1(u,x_4)=\sum_{n=-\infty}^{\infty} \left(\frac{3\frac{u^{\fft32}}{U_K^{\fft12}}\left|\frac{x_4}{R_x}+\pi n\right|}{\frac{u}{U_K}\left(\frac{x_4}{R_x}+\pi n\right)^2+9}
- \frac{3 \sqrt{u}\sqrt{U_K}}{\pi\left|n+\fft12 \right|}\right).
\eea
The other functions we take $N_2=0$ and simply sum on images.  We plot these in figure \ref{plotSumImages}.

In figure \ref{plotSumImages} we have summed many fewer modes for $F_1$ because this function converges very fast (as $1/q^{14}$), and so fewer images are needed.
Further, we can see that one must go to much larger values of $u$ to obtain well separated features for
$\ph_1$.  One could have guessed this from the results of the decompactified case: the characteristic width of the functions $K_3(q)$ (associated with $\ph_1$) is much larger than for the other functions $K_1$ and $K_2$.  However, one can see the shape of the functions approaching the decompactified case: $F_1$ has a single cusp like $-|x_4|$ and $\ph_1$ and $F_2$ have cusps like $|x_4|$ surrounded by two maxima.  We show plots for various $u$ in the next subsection where we consider the Fourier transform of the equations.

\subsubsection*{Fourier decomposition}
We start with
the same equations as the last subsection
\bea
&& \kern-3em 3\pau^2F_1+\frac{12\pau F_1}{u}
+\frac{g_s Q_c \pax^2 F_1}{u^3}
-\frac{54  F_1}{u^2}+\frac{2(Q_c g_s)^{\fft23}\sqrt{3}}{u^2}\sqrt{\frac{u}{(Q_c g_s)^{\fft13}}}\Delta=0 \\
&& \kern-3em 3\pau^2F_2+\frac{12\pau F_2}{u}
+\frac{ g_s Q_c  \pax^2 F_2}{u^3}
-\frac{4(Q_c g_s)^{\fft23}\sqrt{3}}{u^2}\sqrt{\frac{u}{(Q_c g_s)^{\fft13}}}\Delta=0 \\
&& \kern-3em -4 \pau^2 \ph_1 -\fft43\frac{Q_cg_s \pax^2 \ph_1}{u^3}-\frac{2\pau \ph_1}{u}
+\frac{\fft43 (Q_c g_s)^{\fft23}\sqrt{3}}{u^2}\sqrt{\frac{u}{(Q_c g_s)^{\fft13}}}\Delta=0
\eea
and Fourier transform them
\bea
&&\kern-3em 3\pau^2F_{m,1}+\frac{12\pau F_{m,1}}{u}
-\frac{m^2\frac{27 U_K}{4} F_{m,1}}{u^3}
  -\frac{54 F_{m,1}}{u^2}+\frac{9\sqrt{U_K}(1+(-1)^m)}{2\pi u^{\fft32}}=0 \label{def1f0} \\
&&\kern-3em 3\pau^2F_{m,2}+\frac{12\pau F_{m,2}}{u}
-\frac{m^2 \frac{27 U_K}{4}  F_{m,2}}{u^3}
 -\frac{18\sqrt{U_K}(1+(-1)^m)}{2\pi u^{\fft32}}=0 \label{def2f0}\\
&&\kern-3em-4 \pau^2 \ph_{m,1} +\fft43\frac{m^2 \frac{27 U_K}{4} \ph_{m,1}}{u^3}-\frac{2\pau \ph_{m,1}}{u}
+\frac{6 \sqrt{U_K}(1+(-1)^m)}{2\pi u^{\fft32}}=0. \label{dephf0}
\eea
where we have assumed that $x_4$ is periodic with
\be
x_4=x_4+2\pi R_x, \quad R_x^2=\frac{4}{27}\frac{Q_c g_s}{U_K}=\frac{4}{9}\frac{R_{D4}^3}{U_K}
\ee
as in the last section (the reason for this parameterization is explained at the beginning of section 4).

The above equations can be brought to a simple and familiar form by the following change
of coordinates and functions
\bea
F_{m,1}(u)&=& \frac{G_{m,1}\left(\sqrt{\frac{9 m^2 U_K}{u}} \right)}{u^{\fft32}} \nn \\
F_{m,2}(u)&=& \frac{G_{m,2}\left(\sqrt{\frac{9 m^2 U_K}{u}} \right)}{u^{\fft32}} \nn \\
\ph_{m,1}(u)&=& G_{m,3}\left(\sqrt{\frac{9 m^2 U_K}{u}} \right) u^{\fft12} \nn \\
u&=&\frac{9 m^2 U_K}{\rho^2} \nn
\eea
The above differential equations become
\bea
\rho^2 \pap^2 G_{m,1} + \rho \pap G_{m,1} +\left(-\rho^2-9^2\right)G_{m,1}&=&-\frac{972 (1+(-1)^m) m^4 U_K^{\fft52}}{2 \pi \rho^4} \\
\rho^2 \pap^2 G_{m,2} + \rho \pap G_{m,2} +\left(-\rho^2-3^2\right)G_{m,1}&=&\frac{1944 (1+(-1)^m) m^4 U_K^{\fft52}}{2 \pi \rho^4} \\
\pap^2 G_{m,3}- G_{m,3}&=&\frac{6 (1+(-1)^m) U_K^{\fft12}}{2\pi \rho^2}.
\eea
These then are of two forms: modified Bessel equations with a simple monomial source\footnote{Solutions to these equations are known as (modified) Lommel functions.  However they are related to generalized hypergeometric functions.  We opt to use the
notation of hypergeometric series, as these are more general, and perhaps more familiar to the reader.}, and an exponential function with a simple monomial source.  The general solution to the above equations is \cite{GandR}
\bea
\kern-3.5em G_{m,1} &=& C_1 I_9(\rho)+C_2 K_9(\rho)- \frac{972 (1+(-1)^m) m^4 U_K^{\fft52}}{2 \pi \rho^4(-13)(5)}\;_1F_2\left(1;-\fft{11}{2},\fft{7}{2};\frac{\rho^2}{4}\right) \label{F1sln}\\
\kern-3.5em G_{m,2} &=& C_3 I_3(\rho)+C_4 K_3(\rho)+ \frac{1944 (1+(-1)^m) m^4 U_K^{\fft52}}{2 \pi \rho^4(-7)(-1)}\;_1F_2\left(1;-\fft{5}{2},\fft{1}{2};\frac{\rho^2}{4}\right) \label{F2sln} \\
\kern-3.5em G_{m,3} &=& C_5 e^{\rho} + C_6 e^{-\rho} +\frac{3(1+(-1)^m)(e^{\rho}\mbox{Ei}(1,\rho)+e^{-\rho}\Re(\mbox{Ei}(1,-\rho))U_K^{\fft12})}{2\pi} \label{phsln}
\eea
where $I$ and $K$ are the modified Bessel functions, $\;_1F_2$ is the generalized hypergeometric function \footnote{we do in fact mean $\;_1F_2$, not $\;_2F_1$}, and $\mbox{Ei}(a, x)$ is the exponential integral function.
\be
\mbox{Ei}(a, x)=\int_1^{\infty} e^{(-y x)} y^{-a} dy
\ee
and we have used $\Re(\mbox{Ei}(1,-\rho))=\mbox{Ei}(1,-\rho)+\pi i$ to remove the $-\pi i$ associated with going around the branch point at $x=0$.

Here, one may worry about the convergence of the above Fourier decomposition because the
coefficient above depend on $m$ in positive powers.  However, recall that we wish to sum on $m$ for fixed $u$,
not fixed $\rho$.  In fact the factor of $m$ completely cancels out of the above coefficients once returning
to the $u$ coordinate ($\rho^4 \propto m^4/u^2$).  The only $m$ dependence comes about in the arguments of the homogeneous and non homogeneous terms.  Therefore, we may effectively analyze convergence of the Fourier modes as
convergence in the variable $\rho \rightarrow \infty$, as this is the limit to which $m\rightarrow \infty $ corresponds.
The inhomogeneous solution to (\ref{phsln}) is indeed convergent and admits a power series expansion about $\rho=\infty $.  Therefore one does not wish to turn on the growing exponential (this would not converge summing on $m$), and the shrinking exponential is simply negligible.  Hence, we may set $C_5=C_6=0$ for all m and get a convergent series.

The remaining equations, however, deserve some special treatment.  The bessel equation (and the equation for $\;_1F_2$) have an essential singularity at $\rho=\infty$.  Therefore we consider the asymptotics of the above functions for large $\rho$:
\bea
I_{\nu}(\rho)&& \rightarrow \frac{1}{\sqrt{2 \pi \rho}}e^{\rho}    \nn \\
K_{\nu}(\rho)&& \rightarrow \frac{1}{\sqrt{2 \pi \rho}}e^{-\rho}
\eea
and \cite{wolfram}
\bea
\;_1F_2(a_1;b_1,b_2;x)&& \rightarrow   \\
&& \kern-2em \frac{\Gamma(b_1) \Gamma(b_2)}{\sqrt{\pi} \Gamma(a_1)}(-x)^{\fft12(a_1-b_1-b_2)}\cos \left(\frac{\pi}{2}\left(a_1-b_1-b_2+\fft12\right)+2\sqrt{-x}\right) \nn.
\eea
After substituting in $x=\rho^2/2$ and choosing appropriate branches for the square roots ($(-1)^{1/2}=+i$), one finds
\bea
\frac{\;_1F_2(1;-\frac{11}{2},\frac{7}{2};\frac{\rho^2}{4})}{\rho^4}&&\rightarrow \frac{\sqrt{\pi}\sqrt{2}}{2772 \rho^{\fft12}} e^{\rho} \nn \\
\frac{\;_1F_2(1;-\frac{5}{2},\frac{1}{2};\frac{\rho^2}{4})}{\rho^4}&&\rightarrow -\frac{\sqrt{\pi}\sqrt{2}}{60 \rho^{\fft12}} e^{\rho}.
\eea
We therefore use the following combinations
\bea
\kern-3.5em && G_{m,1} = C_2 K_9(\rho)- \frac{972 (1+(-1)^m) m^4 U_K^{\fft52}}{2 \pi (-13)(5)}\left(\frac{\;_1F_2\left(1;-\fft{11}{2},\fft{7}{2};\frac{\rho^2}{4}\right)}{\rho^4}-\frac{\pi I_9(\rho)}{1386}\right) \label{F1slnR}\\
\kern-3.5em && G_{m,2} = C_4 K_3(\rho)+ \frac{1944 (1+(-1)^m) m^4 U_K^{\fft52}}{2 \pi (-7)(-1)}\left(\frac{\;_1F_2\left(1;-\fft{5}{2},\fft{1}{2};\frac{\rho^2}{4}\right)}{\rho^4}+\frac{\pi I_3(\rho)}{30}\right) \label{F2slnR} \\
\kern-3.5em && G_{m,3} = C_6 e^{-\rho} +\frac{3(1+(-1)^m)(e^{\rho}\mbox{Ei}(1,\rho)+e^{-\rho}\Re(\mbox{Ei}(1,-\rho))U_K^{\fft12})}{2\pi}. \label{phslnR}
\eea
It is now a simple matter to replace the definition of $\rho$ above and sum the series.  Although we offer no analytic proof here, the above functions can be seen to converge quickly enough for large $m$.  We note that because the above functions are functions of $m^2/u$, the convergence in $m$ and $u\rightarrow 0$ are connected.  As promised, the Fourier expansion converges more quickly for smaller $u$.

In the last two equations, it should be noted that for large $u$, the particular solution selected dominates over
the remaining inhomogeneous solution.  However, in the first equation, the homogeneous solution dominates.  In the following, we still set $C_2=C_4=C_6=0$.

Solving the $m=0$ case is trivial, but for completeness, we give the solutions for this as well
\bea
&&F_{0,1}=-\frac{12 u^{\fft12}U_K^{\fft12}}{65 \pi}+ C_{0,5}u^3+\frac{C_{0,6}}{u^6} \nn \\
&&F_{0,2}=-\frac{24 u^{\fft12} U_K^{\fft12}}{7 \pi}+\frac{C_{0,3}}{u^3}+C_{0,4} \\
&&\ph_{0,1}=-\frac{3 U_K^{\fft12} \ln(u) u^{\fft12}}{\pi}+\frac{6 u^{\fft12} U_K^{\fft12}}{\pi}+2 C_{0,1} u^{\fft12}+C_{0,2}
\eea
and we again set the unfixed constants above to 0.

\begin{figure}[ht!]
\centering
\subfloat[$\phi_1(u,x_4)/U_K$]{
    \includegraphics[width=0.42\textwidth]{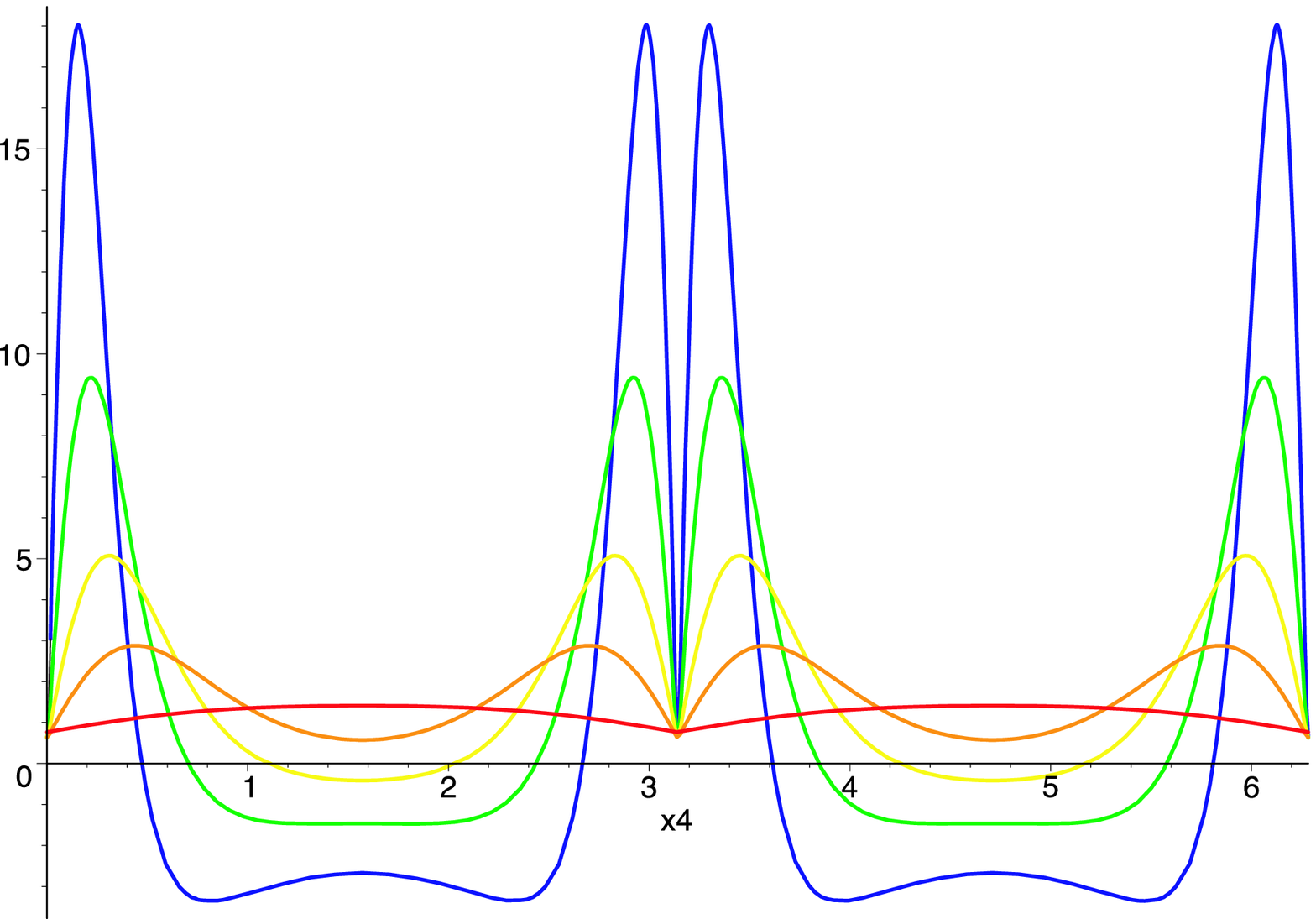}}
\subfloat[$A_1(u,x_4)/U_K$]{
    \includegraphics[width=0.42\textwidth]{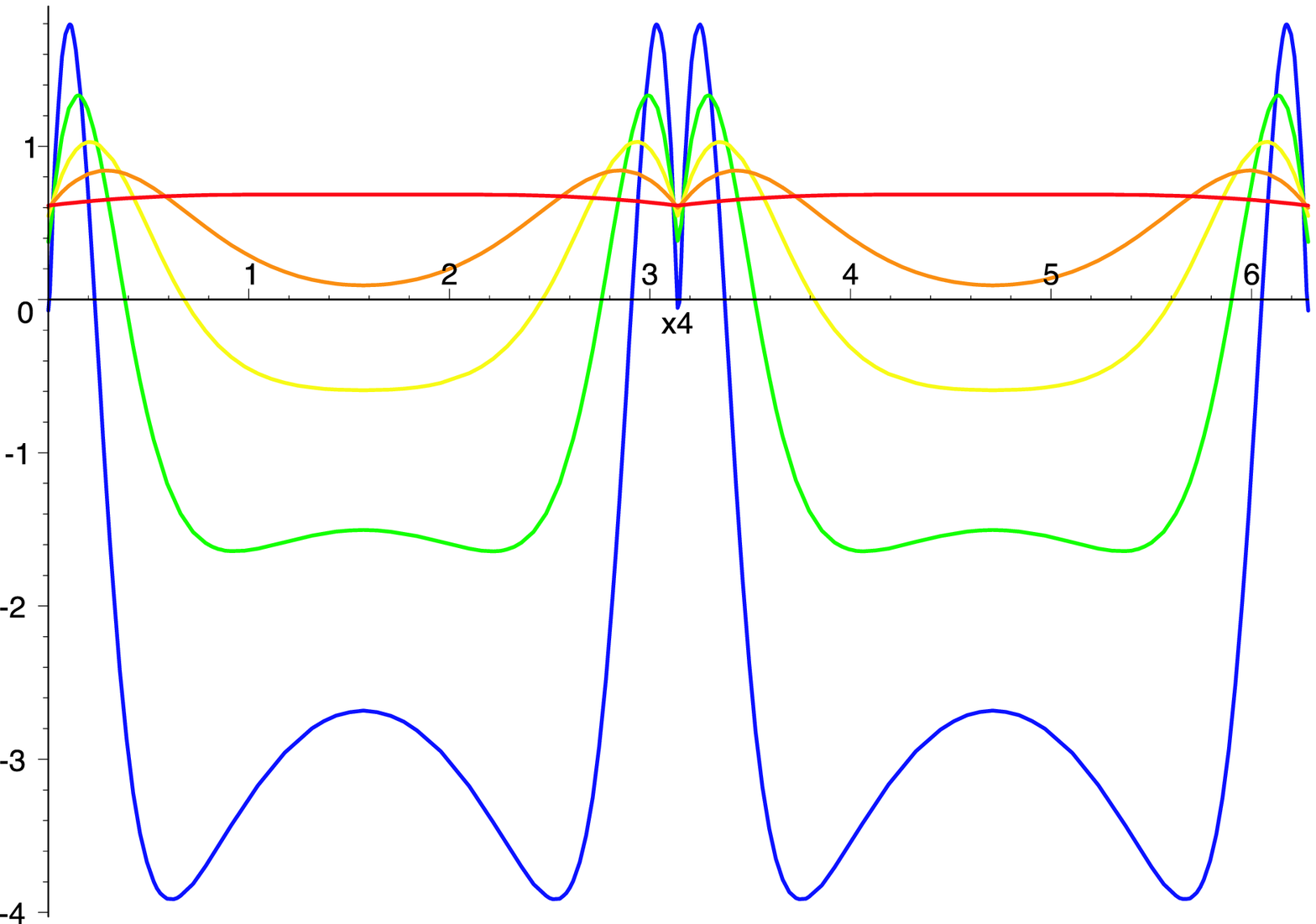}} \\
\subfloat[$B_1(u,x_4)/U_K$]{
    \includegraphics[width=0.42\textwidth]{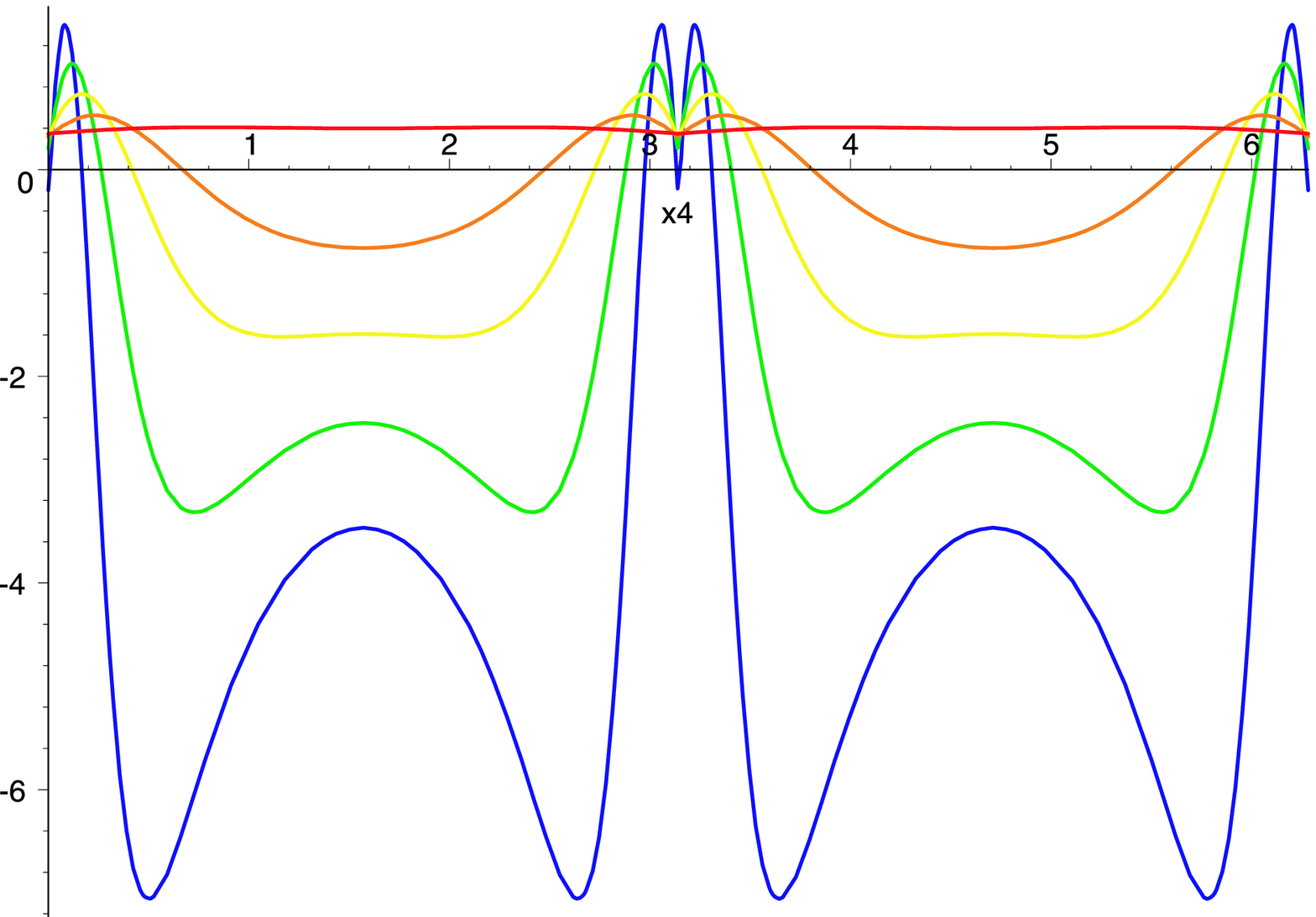}}
\subfloat[$C_1(u,x_4)/U_K$]{
    \includegraphics[width=0.42\textwidth]{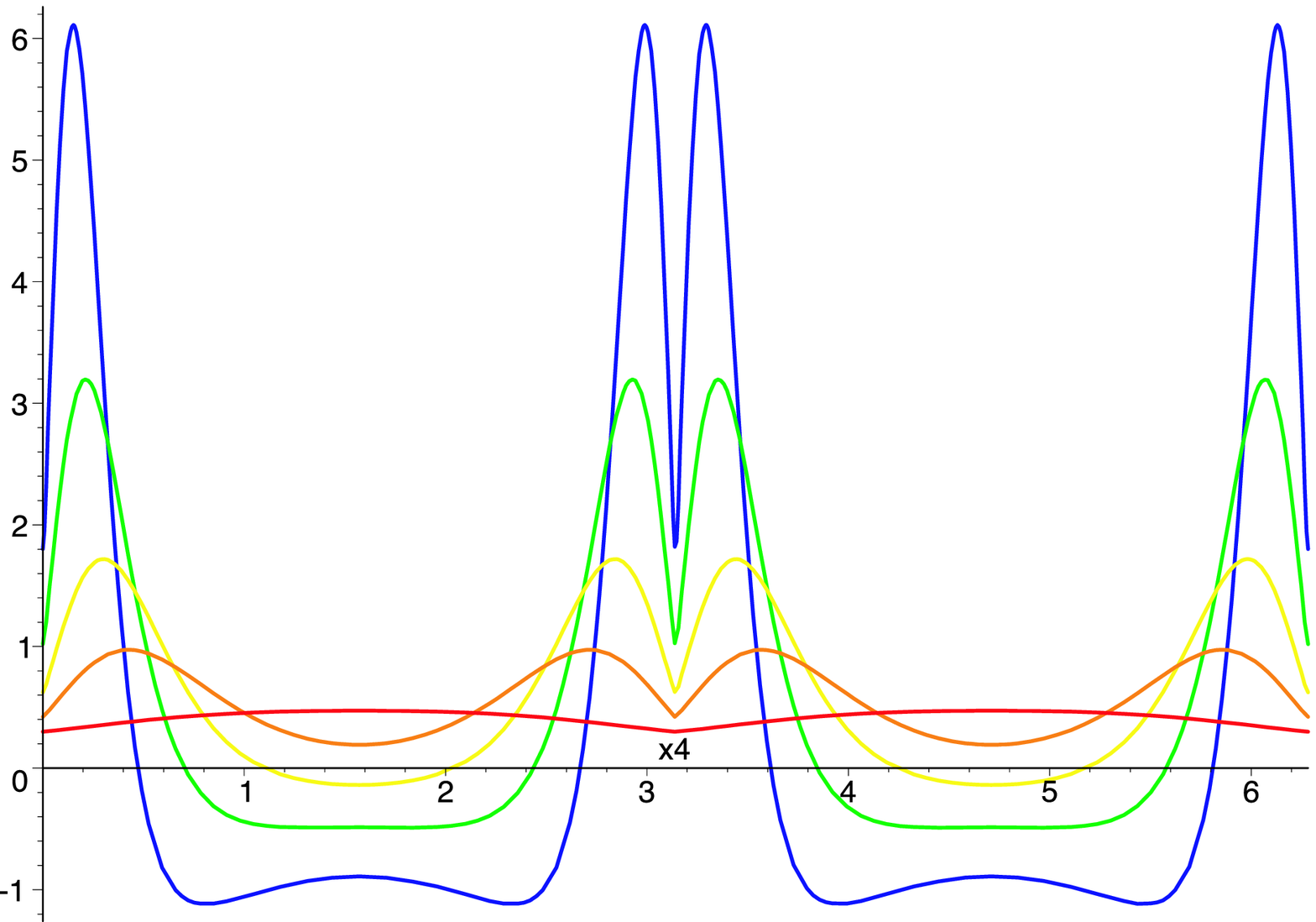}}
\caption{Graphs of $\phi_1/U_K$, $A_1/U_K$, $B_1/U_K=G_1/U_K$, $C_1/U_K$ graphed as a function of $x_4/R_x$.  Each graph represents a different $u$ value of $u=1,5,10,20,40$ in ROYGB order.  All graphs use the first 150 modes ($m=0...150$), of which only even m are non zero.}
\label{PlotForMode}
\end{figure}

We show here the plots of the physical fields $\phi_1,A_1,B_1=G_1,C_1$ for the first $150$ modes in figure \ref{PlotForMode}.
\newpage

\begin{figure}[ht!]
\centering
\subfloat[$F_1(u,x_4)/U_K$]{
    \includegraphics[width=0.3\textwidth,angle=-90]{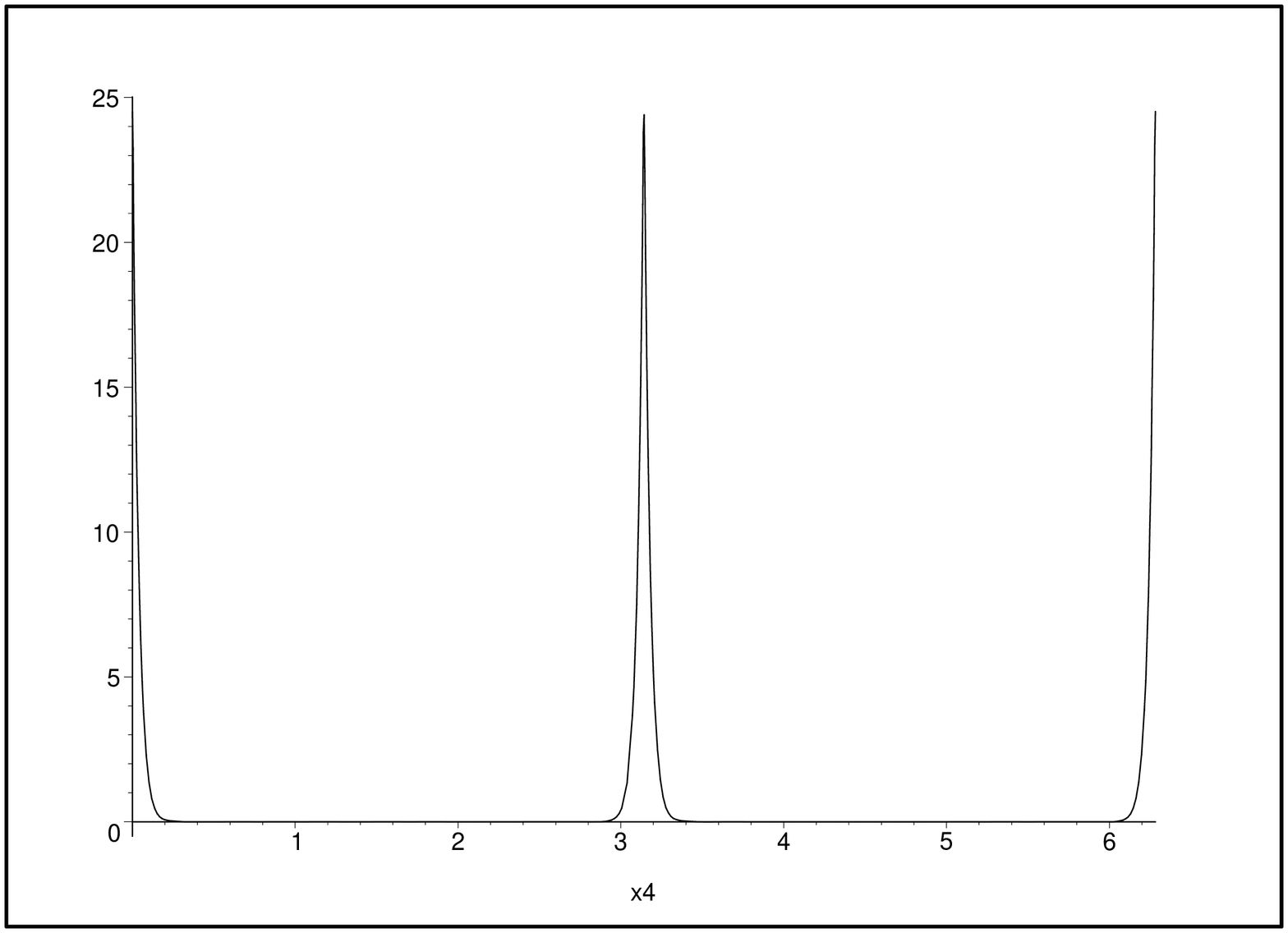}}
\subfloat[$F_2(u,x_4)/U_K$]{
    \includegraphics[width=0.3\textwidth,angle=-90]{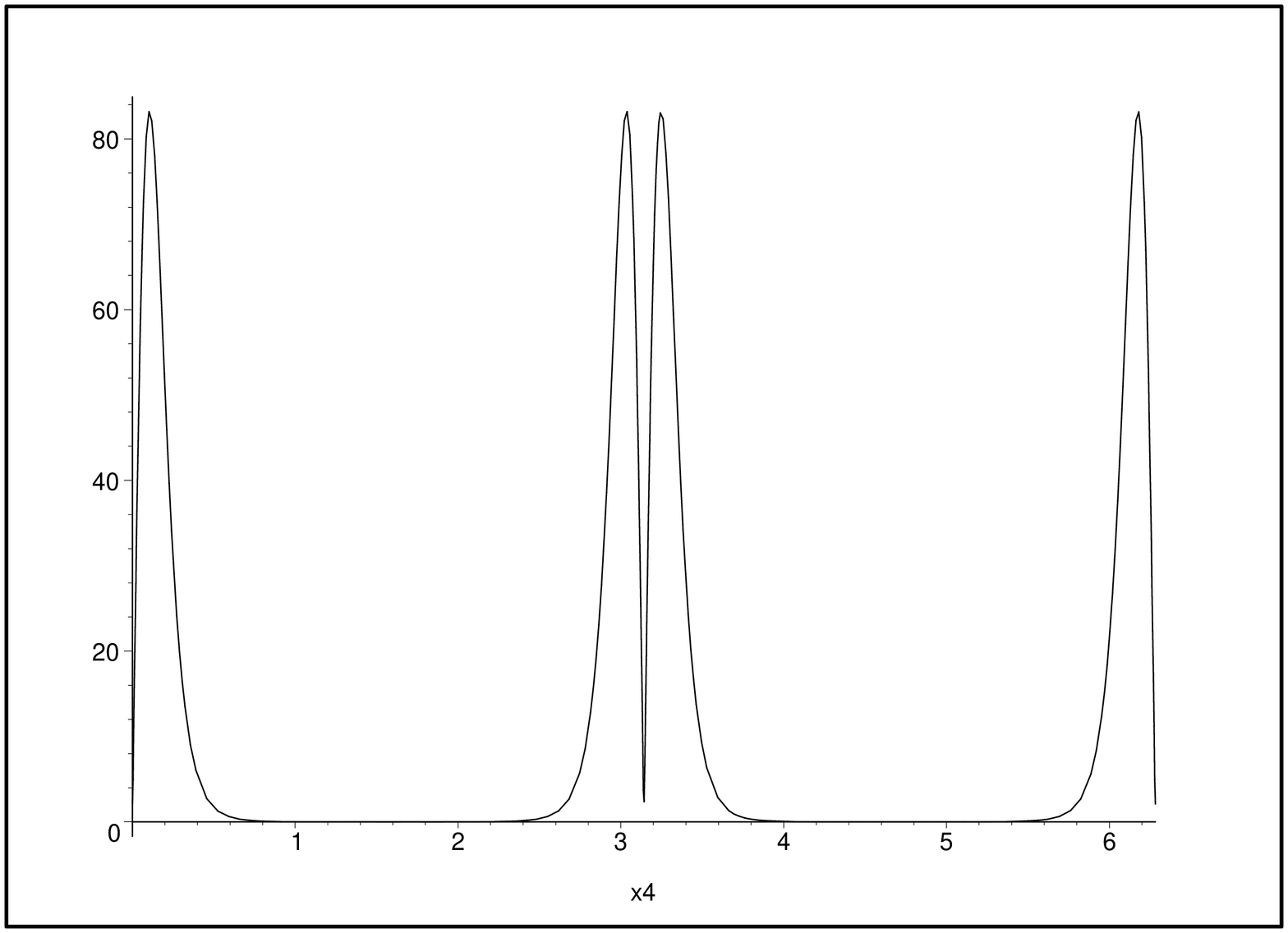}} \\
\subfloat[$\ph_1(u,x_4)/U_K$]{
    \includegraphics[width=0.3\textwidth,angle=-90]{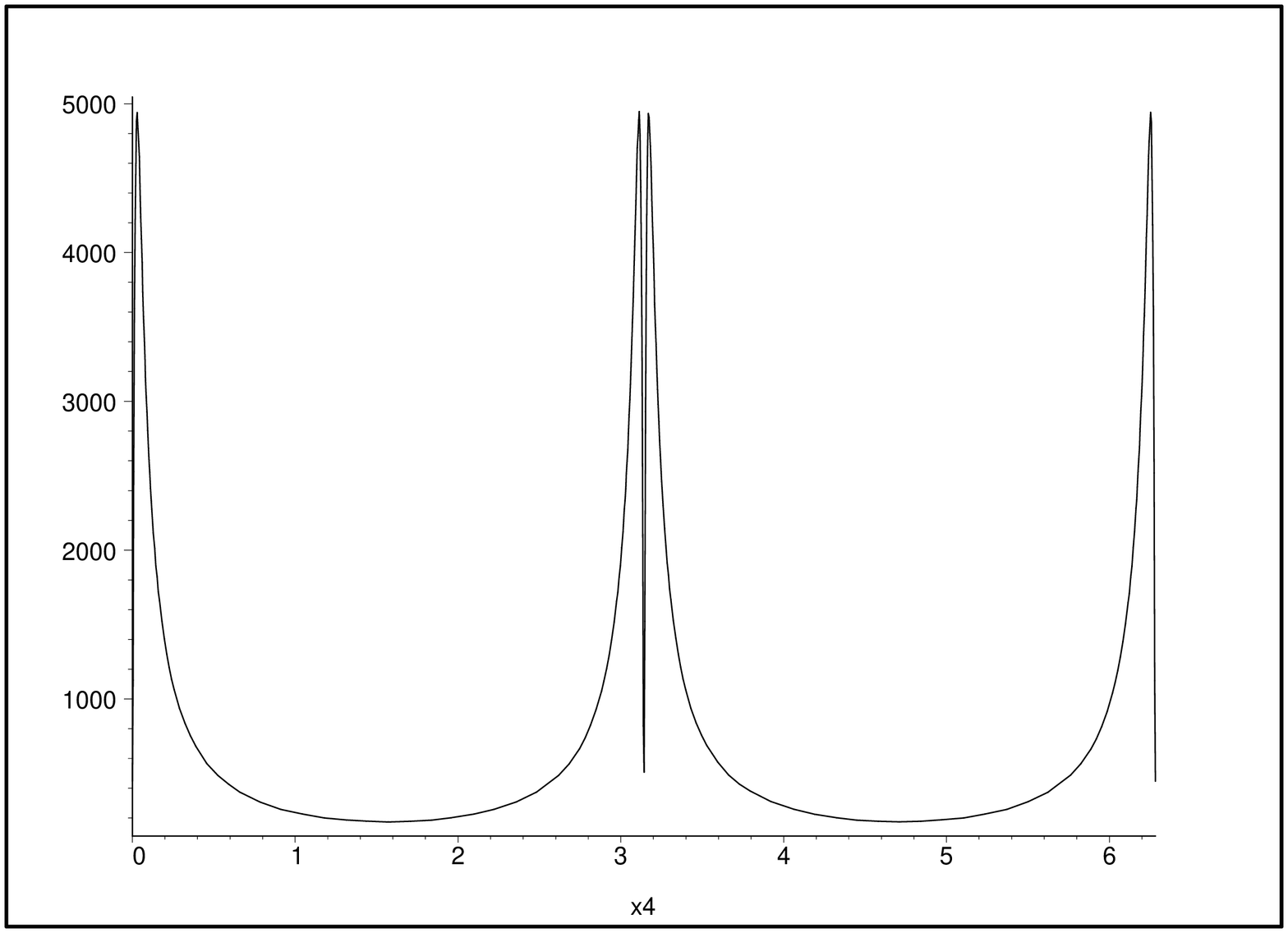}}
\caption{$F_1/U_K$ as a function of $x_4/R_x$ for $u/U_K=100$,
$F_2/U_K$ as a function of $x_4/R_x$ for $u/U_K=100$ and
 $\ph_1/U_K$ as a function of $x_4/R_x$ for $u/U_K=10,000$ (bottom) all graphed using the first 400 modes
($m=0...400$).}
\label{PlotForModeBig}
\end{figure}

We may wish to ask what the long distance behavior is in $u/U_K$.  For this, we plot the long distance
behavior of the independent modes $F_1,F_2,\ph_1$ in figure \ref{PlotForModeBig}.
\newpage

The graphs in figure \ref{PlotForModeBig} are directly comparable to the plots in figure \ref{plotSumImages}, however there may be discrepancies between the two.  For example, in figure \ref{PlotForModeBig} one notes that the function $\ph$ does not quite touch 0 at $x_4=\pi R_x/2$ as it is required to for figure \ref{plotSumImages}.  This is merely a matter of setting the correct zero mode for the homogenous equation in this case.  However, to make the plots match {\it exactly} in all cases, one may, if one wishes, Poisson resum the results from the last subsection, which will fix all constants above in terms of the constants from the last subsection.  We, however, have not done this.

We should note also that the above analysis allows for more generality that summing the images.  In the decompactification case a very specific functional form was taken, to which many homogenous solutions do not conform.  This property is inherited when summing on images.  By the orthogonality and completeness of the trigonometric functions, we are guaranteed to generate all solutions to the original gravitational ansatz when using the Fourier analysis.

%%%%%%%%%%%%%%%%%%%%%%%%%%%%%%%%%%%%%%%%%%%
\subsection{$U_K\neq0$}
%%%%%%%%%%%%%%%%%%%%%%%%%%%%%%%%%%%%%%%%%%%
\subsubsection*{Fourier decomposition}
Fourier decomposing the general equations at the beginning of section 3 gives
\bea
&&3\pau^2F_{m,1}+\frac{3(4u^3-U_K^3)\pau F_{m,1}}{u(u^3-U_K^3)}
-\frac{m^2\frac{27 U_K}{4} u^3 F_{m,1}}{(u^3-U_K^3)^2} \nn \\
&& \quad \quad  -\frac{54 u F_{m,1}}{(u^3-U_K^3)}+\frac{9u\sqrt{U_K}(1+(-1)^m)}{2\pi (u^3-U_K^3)}\sqrt{\frac{u}{\left(1-\frac{U_K^3}{u^3}\right)}}=0 \label{def1f}\\
&& \phantom{x} \nn \\
&& \phantom{x} \nn \\
&& 3\pau^2F_{m,2}+\frac{3(4u^3-U_K^3)\pau F_{m,2}}{u(u^3-U_K^3)}
-\frac{m^2 \frac{27 U_K}{4} u^3 F_2}{(u^3-U_K^3)^2} \nn \\
&& \quad \quad -\frac{18u\sqrt{U_K}(1+(-1)^m)}{2\pi(u^3-U_K^3)}\sqrt{\frac{u}{\left(1-\frac{U_K^3}{u^3}\right)}}=0 \label{def2f}\\
&& \phantom{x} \nn \\
&& \phantom{x} \nn \\
&&-4 \pau^2 \ph_{m,1} -\frac{2\left(u^3-7U_K^3\right)\pau \ph_{m,1}}{u\left(u^3-U_K^3\right)}-\frac{36 u U_K^3 \ph_{m,1}}{\left(u^3-U_K^3\right)^2} +\fft43\frac{m^2u^3 \frac{27 U_K}{4} \ph_{m,1}}{\left(u^3-U_K^3\right)^2}\nn \\
&& \quad \quad+\frac{6 u \sqrt{U_K}(1+(-1)^m)}{2\pi(u^3-U_K^3)}\sqrt{\frac{u}{\left(1-\frac{U_K^3}{u^3}\right)}}=0. \label{dephf}
\eea
where now the periodicity of $x_4$
\be
x_4=x_4+2\pi R_x, \quad R_x^2=\frac{4}{27}\frac{Q_c g_s}{U_K}=\frac{4}{9}\frac{R_{D4}^3}{U_K}
\ee
is dictated by smoothness of the gravitational solution near the tip of the cigar ($u\rightarrow U_K$).  Recall that we will be considering the antipodal embedding $L=\pi R_x$, so that the delta functions are located at $x_4=0,\pi R_x$.

It is clear from the above that these functions are functions only of $u/U_K$ as $U_K$ is the only dimensionful
parameter left.  Under the coordinate change $\hat{u}=\frac{u}{U_K}$ and the redefinition of fields $F_i=\hat{F}_i\times U_K$ and $\ph_1=\tilde{\phi}_1\times U_K$ we find
\bea
&&3\pahu^2\hF_{m,1}+\frac{3(4\hu^3-1)\pahu \hF_{m,1}}{\hu(\hu^3-1)}
-\frac{27}{4}\frac{m^2 \hu^3 \hF_{m,1}}{(\hu^3-1)^2} \nn \\
&& \quad \quad  -\frac{54 \hu \hF_{m,1}}{(\hu^3-1)}+\frac{9\hu^3(1+(-1)^m)}{2\pi (\hu^3-1)^{\fft32}}=0 \label{def1ff}\\
&& \phantom{x} \nn \\
&& \phantom{x} \nn \\
&& 3\pahu^2\hF_{m,2}+\frac{3(4\hu^3-1)\pahu \hF_{m,2}}{\hu(\hu^3- 1)}
-\frac{27 }{4}\frac{m^2  \hu^3 \hF_{m,2}}{(\hu^3-1)^2} \nn \\
&& \quad \quad -\frac{18\hu^3(1+(-1)^m)}{2\pi (\hu^3-1)^{\fft32}}=0 \label{def2ff}\\
&& \phantom{x} \nn \\
&& \phantom{x} \nn \\
&&-4 \pahu^2 \hhp_{m,1} -\frac{2\left(\hu^3-7\right)\pahu \hhp_{m,1}}{\hu\left(\hu^3-1\right)}-\frac{36 \hu \hhp_{m,1}}{\left(\hu^3-1\right)^2} +9\frac{m^2\hu^3 \hhp_{m,1}}{\left(\hu^3-1\right)^2}\nn \\
&& \quad \quad+\frac{6 \hu^3(1+(-1)^m) }{2\pi (\hu^3-1)^{\fft32}}=0. \label{dephff}
\eea

The above differential equations are difficult to solve, even excluding the inhomogeneous piece.  The homogeneous parts of the equations can be seen to have 5 regular singular points at $\hat u=0,\omega^0,\omega^1,\omega^2,\infty$ where
$\omega=-\fft12+\fft{\sqrt{3}}{2}i$ is a third root of unity.  However, as all of the singularities are regular, one may go about finding a Laurent series expansion about any given point (using standard textbook techniques). Rather than doing this in the Fourier basis, we find that it is easier to leave the functions of $x_4$ intact, and expand the function about a specific $u$ value, and then solve for the functions of $x_4$ multiplying each power individually.

\subsubsection*{$u\rightarrow U_K$ expansion}
First, for expanding around the point $u=U_K$, and to make contact with much of the literature, we change
to the coordinate $z$ defined by
\be
u^3=U_K^3+U_Kz^2. \label{defz}
\ee
In these coordinates, the (non algebraic) equations read
\bea
&&\frac{Q_c g_s(U_K^2+z^2)\pax^2F_1}{(U_K^3+U_Kz^2)^{\fft13}z^4U_K}+\fft{27}{4}\frac{(U_K^2+z^2)\paz^2 F_1}{U_Kz^2} +\fft{27}{4}\frac{(U_K^2+3z^2)\paz F_1}{U_K z^3} \nn \\
&& \quad +\frac{54 F_1}{U_Kz^2}-\frac{2(3Q_cg_s)^{\fft12}\sqrt{(U_K(U_K^2+z^2)^4)^{\fft13}}\Delta}{U_Kz^3}=0 \\
\phantom{1} \nn \\
\phantom{1} \nn \\
&&\frac{Q_c g_s(U_K^2+z^2)\pax^2F_2}{(U_K^3+U_Kz^2)^{\fft13}z^4U_K}+\fft{27}{4}\frac{(U_K^2+z^2)\paz^2 F_2}{U_Kz^2} +\fft{27}{4}\frac{(U_K^2+3z^2)\paz F_2}{U_K z^3} \nn \\
&& \quad -\frac{4(3Q_cg_s)^{\fft12}\sqrt{(U_K(U_K^2+z^2)^4)^{\fft13}}\Delta}{U_Kz^3}=0 \\
\phantom{1} \nn \\
\phantom{1} \nn \\
&&-\fft43\frac{Q_c g_s(U_K^2+z^2)\pax^2\ph_1}{(U_K^3+U_Kz^2)^{\fft13}z^4U_K}-\fft{9}{1}\frac{(U_K^2+z^2)\paz^2 \ph_1}{U_Kz^2} -3\frac{(2z^2-9U_K^2)\paz \ph_1}{U_K z^3} \nn \\
&& \quad -\frac{36 U_K \ph_1}{z^4}+\frac{\fft{4}{3}(3Q_cg_s)^{\fft12}\sqrt{(U_K(U_K^2+z^2)^4)^{\fft13}}\Delta}{U_Kz^3}=0 \\
\phantom{1} \nn \\
\phantom{1} \nn \\
&&B_1=G_1=-\fft15 F_1+\fft15F_2-\fft35\frac{\paz\ph_1 (U_K^2+z^2)}{z}+\fft{1}{10}\frac{\ph_1(6U_K^2-z^2)}{z^2} \nn \\
&& \phantom{1}
\eea
after a bit of cleaning up.

Now, to expand about $z=0$, we expand the functions multiplying each differential operator that acts on the fields $F_i, \ph$ above.
\bea
&&\kern-3.4em \frac{Q_c g_s\pax^2F_1}{z^4}+\fft{27}{4}\frac{U_K\paz^2 F_1}{z^2} +\fft{27}{4}\frac{U_K\paz F_1}{z^3} -\frac{54 F_1}{U_Kz^2}+\frac{2(3Q_cg_s)^{\fft12}\sqrt{U_K}\Delta}{z^3}=0 \\
\phantom{1} \nn \\
&&\kern-3.4em \frac{Q_c g_s\pax^2F_2}{z^4}+\fft{27}{4}\frac{U_K\paz^2 F_2}{z^2} +\fft{27}{4}\frac{U_K\paz F_2}{z^3} -\frac{4(3Q_cg_s)^{\fft12}\sqrt{U_K}\Delta}{z^3}=0 \\
\phantom{1} \nn \\
&&\kern-4.0em -\fft43\frac{Q_c g_s \pax^2\ph_1}{z^4}-\frac{9U_K\paz^2 \ph_1}{z^2} +\frac{27U_K^2\paz \ph_1}{ z^3}
-\frac{36 U_K \ph_1}{z^4}+\frac{\fft{4}{3}(3Q_cg_s)^{\fft12}\sqrt{U_K}\Delta}{z^3}=0 \\
\phantom{1} \nn \\
&&\kern-3.4em B_1=G_1=-\fft15 F_1+\fft15F_2-\fft35\frac{U_K^2\paz\ph_1 }{z}+\fft{3}{5}\frac{U_K^2\ph_1}{z^2}
\eea
We will now assume that all of the Fourier modes of the above fields have expansions about $z=0$ with a finite number of negative powers.  This allows us to simply count powers in $z$ and neglect any terms that are not
of leading order.  For example, this allows us to drop the term $F_1/z^2$ above.  The rest of the equation is homogeneous in $z$ only if $F_1\sim z$ as $z\rightarrow 0$.  Likewise, we conclude that to leading order $F_i=z L_i(x_4)$ and $\ph_1=z L_3(x_4)$.  To leading order in $z$, the above equations become
\bea
&&\pax \pax L_1(x4)+\frac{27U_K}{4Q_c g_s} L_1(x4)+2\sqrt{\frac{3U_K}{Q_c g_s}}\Delta=0\\
&&\pax \pax L_2(x4)+\frac{27U_K}{4Q_c g_s} L_2(x4)-4\sqrt{\frac{3U_K}{Q_c g_s}}\Delta=0\\
&&\pax \pax L_3(x4)+\frac{27U_K}{4Q_c g_s} L_3(x4)-\sqrt{\frac{3U_K}{Q_c g_s}}\Delta=0\\
&&B_1=G_1=z\fft15\left(L_2-L_1\right).
\eea
These are easily solved.
\bea
L_1(x_4)&=& -\fft43 \Big|\sin{\left(\frac{x_4}{R_x}\right)}\Big| \\
L_2(x_4)&=& +\fft83 \Big|\sin{\left(\frac{x_4}{R_x}\right)}\Big| \\
L_3(x_4)&=& +\fft23 \Big|\sin{\left(\frac{x_4}{R_x}\right)}\Big| \\
R_x^2&=&\frac{4Q_c g_s}{27U_K}
\eea
where $R_x$ is the periodicity of $x_4\equiv x_4 +2\pi R_x$ defined before.
Note that this is exactly the kind of behavior one would expect because $z|\sin{(x_4/R_x)}|=|y_1|$
when switching to ``cartesian'' $y_2=z \cos{(x_4/R_x)}, y_1=z \sin{(x_4/R_x)}$ coordinates, and
is similar qualitatively to a D8 in flat space.

One may continue this process and in fact get the above functions to the next order.  From the general
differential equation in $z$, one may easily read that the expansion will be in terms of
odd powers of $z$.  This is because the functions being expanded are all $(U_K^2+z^2)^n$, and so coefficients
of even (odd) powers only mix with coefficients of other even (odd) powers.  The next term, therefore, should be of order $z^3$.  We solve the resulting equations, and find
\bea
F_1(z,x_4)&=&-\fft23 z \sin{\left(\frac{|x_4|}{R_x}\right)}
-\frac{17}{108}\frac{z^3\left(3\sin{\left(\frac{|x_4|}{R_x}\right)}
-\sin{\left(\frac{3|x_4|}{R_x}\right)}\right)}{U_K^2} +\mathcal{O}(z^5)\nn \\
F_2(z,x_4)&=&2\times\fft23 z \sin{\left(\frac{|x_4|}{R_x}\right)}
-\frac{14}{108}\frac{z^3\left(3\sin{\left(\frac{|x_4|}{R_x}\right)}
-\sin{\left(\frac{3|x_4|}{R_x}\right)}\right)}{U_K^2}+\mathcal{O}(z^5) \nn \\
\ph_1(z,x4)&=&\fft13z \sin{\left(\frac{|x_4|}{R_x}\right)}+\mathcal{O}(z^5).
\eea
One may analyze the above functions for the various length scales of these cusps.  One finds that they slope is directly limited by $d\phi_i/ds= d\phi_i/dy_1(U_K/R_{D4})^{3/4}\sim (U_K/R_{D4})^{3/4}Q_f\ll 1/\ell_s$.  This gives $(g_5^4 T_{st})^{1/2}N_f\propto (T_{st}/M_{gb}^2)^{1/2}\lambda_4 N_f/N_c\ll1$.

Further, the onset of new ``features'' is at $z^2=U_K^2$, which we require to be a large physical length: $U_K^2 (R_{D4}/U_K)^{3/2}\gg \ell_s^2$, which is equivalent to the original condition $\lambda_5 \gg R_x$ given for the supergravity limit.  Hence, we trust the supergravity approximation to describe the features near $u\rightarrow U_K$.

%%%%%%%%%%%%%%%%%%%%%%%%%%%%%%%%%%%%%%%%%%
\section{Analysis of equations: Stability}
%%%%%%%%%%%%%%%%%%%%%%%%%%%%%%%%%%%%%%%%%%

A few words are in order to explain how we will address the issue of stability.
We will not solve the eigenvalue problem, numerically or otherwise, to establish the
four dimensional masses for fluctuations as being positive definite with a mass gap.
Instead, we will simply show that to quadratic order in fluctuations, all actions are of
the form (up to gauge)
\bea
&&\int d^5 \xi \Bigg[-\mathcal{F}_0(\xi;Q_f)-\mathcal{F}_1(\xi;Q_f)\left(-(\partial_t M)^2+\sum_{i=1}^3(\partial_{x_i} M)^2\right)\nn \\
&&\quad -\mathcal{F}_2(\xi;Q_f)(\partial_{\xi^5} M)^2-\mathcal{F}_3(\xi;Q_f)M^2\Bigg]
\eea
where $M$ is a field describing the fluctuation, and
\be
\mathcal{F}_i(\xi;Q_f=0) > 0.
\ee
This last statement is important, because it implies that anywhere the perturbative analysis is valid
that (to linear order in $Q_f$)
\be
\mathcal{F}_i(\xi;Q_f)=
\mathcal{F}_i(\xi;Q_f=0)\left(1+\frac{Q_f}{\mathcal{F}_i}\left(\partial_{Q_f}\mathcal{F}_i\right)|_{Q_f=0} \right)>0.
\ee
The inequality holds because if the term added to the $1$ must be small.  If it were not small, we would be forced to go beyond linear order in $Q_f$, and hence the perturbative approach would no longer be trusted.  Therefore it does not change the sign of any of the functions where the perturbative analysis is valid.  Hence, the action that we have written had positive definite hamiltonian
\bea
&&\int d^5 \xi \Bigg[\mathcal{F}_0(\xi;Q_f)+\mathcal{F}_1(\xi;Q_f)\left((\partial_t M)^2+\sum_{i=1}^3(\partial_{x_i} M)^2\right)\nn \\
&&\quad +\mathcal{F}_2(\xi;Q_f)(\partial_{\xi^5} M)^2+\mathcal{F}_3(\xi;Q_f)M^2\Bigg]
\eea
simply because it is a sum of squares.

For this reason, what seems to be important is the presence of a cancelation between the DBI and CS action at the order  $|M|$ (we expect such terms from the solutions above).

Of course another interesting question is what happens outside of the regime of validity of the perturbative approach. This, however, is out of the scope of our present investigation, although we hope to address this in some future work.

%%%%%%%%%%%%%%%%%%%%
\subsection{DBI and CS equations of motion: $x_4=0$ ($\pi R_x)$ solution}
%%%%%%%%%%%%%%%%%%%%

Of course to expand an action, we must expand about some {\it solution} to the equations of motion.  For this reason, we briefly outline (and give more detail in appendix C) why $x_4=0$ ($x_4=\pi R_x)$ is still a solution to the equations of motion.
One may address this simply by looking at the gauge invariant information in the equations of motion: namely the cusps.  The cusps in $\ph_1,A_1,G_1$ and $C_1$ cannot be removed
with a coordinate transformations, as this would introduce delta functions into $B_1$.  In appendix C, we show that
the cusp in $B_1$, which is pure gauge, does not enter.  We can read off the behavior around the cusps \footnote{at $x_4=0$, similar conditions apply at $x_4=\pi R_x$}
to be
\bea
\ph_1&=&f_{\ph_1}(u)+\fft12 \frac{\sqrt{3} u^{\fft32} \left(1-\frac{U_K^3}{u^3}\right)^{\fft12}|x_4|}{(Q_c g_s)^{\fft12}} +\mathcal{O}(x_4^2)\nn \\
F_1&=&f_{1}(u) - \fft122\frac{\sqrt{3} u^{\fft32} \left(1-\frac{U_K^3}{u^3}\right)^{\fft12}|x_4|}{(Q_c g_s)^{\fft12}} +\mathcal{O}(x_4^2)\label{x4cusp} \\
F_2&=&f_{2}(u) +\fft12 4\frac{\sqrt{3} u^{\fft32} \left(1-\frac{U_K^3}{u^3}\right)^{\fft12}|x_4|}{(Q_c g_s)^{\fft12}} +\mathcal{O}(x_4^2) \nn
\eea
simply by comparing the delta function source terms and the coefficient (a function of $u$) of the $\pax^2$ term.
The next order contributions are of order $x_4^2$ because we require the functions be even about $x_4=0$ to obey
the $\mathbb{Z}_2$ symmetry of the problem.

Now we need to find the equations of motion for the embedding functions $X^{\mu}$ resulting from the action
\be
-\frac{g_s S_B}{Q_f}=\int d^{9}\xi e^{-\phi} \sqrt{-g_p} + \int A_{9}.
\ee
In the appendix, we show that the only interesting equation of motion is the one
for $X^4$, and we find that this becomes
\bea
&& -\sqrt{-g_p}\frac{u^{-\fft34}}{gs R_{D4}^{-\fft34}}\partial_{x_4}\left(
\frac{u^{\fft32}\left(1-\frac{U_K^3}{u^3}\right)^{\fft12}|x_4|}{(R_{D4})^{\fft32}}+\mathcal{O}(x_4^2)\right) \nn \\
&& \quad-\pm\partial_{x_4}\left(
\sqrt{-g_p}|x_4|\frac{1}{g_s} \frac{u^{\fft34}\left(1-\frac{U_K^3}{u^3}\right)^{\fft12}}{R_{D4}^{\fft34}}+\mathcal{O}(x_4^2)\right) \nn \\
&=&-\sqrt{-g_p}\frac{u^{\fft34}\left(1-\frac{U_K^3}{u^3}\right)^{\fft12}}{R_{D4}^{\fft34}g_s}\partial_{x_4}\left(|x_4|\pm |x_4|\right)
\eea
where in the third line, we ignore the higher order in $x_4$ corrections, as we will evaluate the
derivative at $x_4=0$.  Above we have also switched back to the more familiar $g_s Q_c=3R_{D4}^3$ notation.

The term with the $\pm$ comes from the CS term, and may vanish for the $-$ sign choice above. We interpret this
as putting a brane next to the backreacted branes.  This tells that the
equations of motion are satisfied for a brane placed directly on top of the other branes.  If instead
we had put an anti-brane, we would have found a constant force type potential, and we take that
this is a solution too (although we expect an open string tachyon for small enough distances: our actions
do not contain terms for strings ending on different branes).

Further, we should note that the above is a gauge independent statement.  The cusps in the functions
$A_1,C_1,G_1,A_{(9)}$ are independent of the gauge choice \footnote{in any sense: small coordinate transformation or shifts of $A_{(9)}$ by a infinitely differentiable globally exact form}, and these were the only functions that contribute above (see appendix C).
Therefore, the leading $|x_4|$ dependence is unaffected
by small coordinate transformations.  For this statement, it is important that $B$ does not appear: it's values (cusps
and all) are gauge dependent, while the other functions are determined in a gauge covariant way.  Recall that while $\ph_1$ is gauge dependent, it's cusp behavior is not: one may not remove any part of the cusp
without introducing unwanted delta functions into $B_1$.

This gives that to lowest order, original embedding solution is still a solution to the equations of motion.  To truly
consider the stability, however, we would like to know whether this extremum of the action is a maximum or a minimum.  For this we will need to investigate the second order action about this point, and this
will involve second derivatives in $x_4$, rather than just first derivatives.  Hence the even functions
that we were able to ignore in the above discussion will enter.

%%%%%%%%%%%%%%%%%%%%%%%%%%%%%%%%%%%%%%%%%%
\subsection{$U_K=0$ decompactification limit}
%%%%%%%%%%%%%%%%%%%%%%%%%%%%%%%%%%%%%%%%%%

We begin with the solutions for the decompactified case in the last section
\bea
F_i(u,x_4)&=&u K_i(q) \quad q=\frac{x_4 \sqrt{u}}{(Q_c g_s)^{\fft12}} \nn \\
\ph_1(u,x4)&=& u K_3(q),
\eea
where $K_i$ are given in equation (\ref{Ks}).
We will want to construct the second order action in $x_4$ and so we will need $K_i$
to second order in $x_4\propto q$.  We expand the $K_i$ to obtain $F_i$ and $\ph_1$ and find
\bea
F_1(u,x_4)&=&u K_1(q)=\frac{256}{1001}u-\frac{\sqrt{3}u^{\fft32}|x_4|}{(Q_c g_s)^{\fft12}}+\frac{768 u^2 x_4^2}{143 Q_c g_s} \nn \\
F_2(u,x_4)&=&u K_2(q)=-N_2 \frac{\sqrt{3}}{9}u+ \frac{2\sqrt{3}u^{\fft32}|x_4|}{(Q_c g_s)^{\fft12}}+N_2\frac{2\sqrt{3}u^2 x_4^2}{3 Q_c g_s} \\
\ph_1(u,x4)&=& u K_3(q)=N_3\frac{u}{4}+\fft12\frac{\sqrt{3} u^{\fft32}|x_4|}{(Q_c g_s)^{\fft12}}\nn-N_3\frac{3u^2 x_4^2}{16 Q_c g_s}
\eea
where we have dropped order $\mathcal{O}(x_4^3)$ and higher terms.

We are now able examine stability of the decompactified limit by examining the second order action.  We start by writing
the pullback metric as a function of $X^4(x^{\mu},u)$
\be
ds_p^2=e^{2A}\left(\eta_{\mu \nu}dx^{\mu}dx^{\nu}\right)+e^{2G}du^2+e^{2C}d\Omega_4^2+e^{2B}\left(\partial_\mu X^4+\pau X^4 du\right)^2.
\ee
We need the determinant of this metric to second order in $X^4$, however it is easier not to expand the above
metric completely, and simply realize that
\be
\sqrt{-g_p}=\sqrt{-g_{p0}}\left(1+\fft12 g_{p0}^{a b}h_{a b}\right)+ \mathcal{O}((X^4)^4)
\ee
where $g_{p0}$ is constructed by dropping the last term in $ds_p^2$, and $h_{ab}$ is the symmetric tensor
defined by the last term in $ds_p^2$.  Evaluating this, we find
\bea
\kern-2em -\int d^9\xi e^{-\phi}\sqrt{-g_p}=&& \\
&&\kern-10em -\int d^9\xi e^{-\phi+4A+4C+G}\left(1+e^{2(B-A)}\fft12 \eta^{\mu \nu}\partial_{\mu}X^4 \partial_{\nu} X^4+e^{2(B-G)}\fft12\pau X^4 \pau X^4\right) \nn
\eea
again, dropping order $(X^4)^3$ and higher.  We now take the expansion of the functions $A=A_0+Q_f A_1,\cdots$
and evaluate to zeroth and first order in $Q_f$, using the above expansions about $X^4=0$, and keeping
only those terms second order in $X^4$ or lower.
\bea
-\int d^9\xi e^{-\phi}\sqrt{-g_p}=&& \nn \\
&&\kern-8em -\int d^9\xi \frac{u^4\left(\frac{Q_c g_s}{3}\right)^{\fft12}}{g_su^{\fft32}}
\Bigg[1-Q_f\left(\frac{512}{5005}+N_2\frac{4\sqrt{3}}{45}+N_3\frac{9}{20}\right)u \nn \\
&&\kern-6em+Q_f\frac{\sqrt{3}u^{\fft32}|X^4|}{\sqrt{Q_c g_s}}
+Q_f\left(-\frac{1536}{715}+N_2\frac{8\sqrt{3}}{15}+N_3\frac{33}{80}\right)\frac{u^2X_4^2}{Q_c g_s}  \\
&&\kern-6em+\left(1-Q_f\left(\frac{512}{5005}+N_2\frac{\sqrt{3}}{9}+N_3\frac{11}{20}\right)u\right)\fft12\eta^{\mu \nu}\partial_{\mu}X^4 \partial_{\nu} X^4\nn \\
&&\kern-6em+\frac{3u^3}{Q_c g_s}\left(1-Q_f\left(\frac{512}{5005}+N_2\frac{4\sqrt{3}}{45}+N_3\frac{9}{20}\right)u\right)\fft12\pau X^4 \pau X^4\Bigg]. \nn
\eea
To this we must add the term coming from the RR coupling.  To do so, we recognize that
\be
A_{\mu_i|x_4}=\mp \sqrt{-g_p}\frac{Q_f}{g_s}e^{B}|x_4| \label{A9UK0}
\ee
satisfies the equations of motion for the nine form potential.  We use the notation $|x_4$ to mean that the index for $x_4$ has been omitted.  The factor of $-g_p$ in (\ref{A9UK0})
is constructed using only the {\it zeroth} order in $Q_f$ metric pulled back.  Further, this $-g_p$ only has corrections of order $\mathcal{O}((X^4)^2)$,
and so we may ignore them because of the $|x_4|$ already multiplying $\sqrt{-g_p}$.  Hence, to the order that we are working,
\be
-\int A_9=-\int d^9\xi (\mp)u^4\frac{Q_f}{g_s} |X^4|.
\ee
Thus, for the correct orientation of the probe brane, the $|X^4|$ term in the action completely cancels \footnote{We interpret this as a brane next to the backreacted brane(s), rather than an anti brane next to the backreacted brane(s).}, and the total action becomes
\bea
-\int&&\!\!\!\! d^9\xi e^{-\phi}\sqrt{-g_p}= \nn \\
&&\quad -\int d^9\xi \frac{u^4\left(\frac{Q_c g_s}{3}\right)^{\fft12}}{g_su^{\fft32}}\Bigg[1-Q_f\left(\frac{512}{5005}+N_2\frac{4\sqrt{3}}{45}+N_3\frac{9}{20}\right)u \nn \\
&&\quad+Q_f\left(-\frac{1536}{715}+N_2\frac{8\sqrt{3}}{15}+N_3\frac{33}{80}\right)
\frac{u^2X_4^2}{Q_c g_s}  \\
&&\quad+\left(1-Q_f\left(\frac{512}{5005}+N_2\frac{\sqrt{3}}{9}+N_3\frac{11}{20}\right)u\right)\fft12\eta^{\mu \nu}\partial_{\mu}X^4 \partial_{\nu} X^4\nn \\
&&\quad+\frac{3u^3}{Q_c g_s}\left(1-Q_f\left(\frac{512}{5005}+N_2\frac{4\sqrt{3}}{45}+N_3\frac{9}{20}\right)u\right)\fft12\pau X^4 \pau X^4\Bigg]. \nn
\eea

Although we can at this point find equations of motion for the above action, and proceed with the analysis directly, we find it convenient to manipulate the above equation a bit more.  For this, we note that we can redefine
the $u$ coordinate as well as the field $X^4$.  We find it convenient to do the following transformation
\bea
u&=&\hat{u}+ Q_f \lambda_1 \hat{u}^2  \nn \\
du&=&(1 + 2 Q_f \lambda_1 \hat{u})d\hat{u} \nn \\
X^4(x^{\mu},\hat{u})&=&\left(1+\fft12 Q_f \lambda_2 \hat{u}\right) \hat{X}^4(x^{\mu},\hat{u})
\eea
(we do the $u$ coordinate change first, and then the $X$ transformation) and then for ease of notation we simply drop the $\hat{\;}$ from the above.  Again, we may only keep order $Q_f$ or lower in the above expansion.  After doing this, we will introduce
terms of the form $Q_f \lambda_2 f(u) X^4 \pau X^4=\fft12 Q_f \lambda_2 f(u) \pau((X^4)^2)$ which we integrate by parts. This affects the coefficient of $(X^4)^2$.  After doing so, we find that the new action
is
\bea
&&-\int d^9\xi e^{-\phi}\sqrt{-g_p}= \nn \\
&&\quad -\int d^9\xi \frac{u^4\left(\frac{Q_c g_s}{3}\right)^{\fft12}}{g_su^{\fft32}}\Bigg[1-Q_f\left(\frac{512}{5005}+N_2\frac{4\sqrt{3}}{45}+N_3\frac{9}{20}-\fft92 \lambda_1\right)u \nn \\
&&\quad+Q_f\left(-\frac{1536}{715}+N_2\frac{8\sqrt{3}}{15}+N_3\frac{33}{80}-\fft{11}{8}\lambda_2\right)
\frac{u^2X_4^2}{Q_c g_s}  \\
&&\quad+\left(1-Q_f\left(\frac{512}{5005}+N_2\frac{\sqrt{3}}{9}+N_3\frac{11}{20}-\fft92\lambda_1-\lambda_2\right)
u\right)\fft12\eta^{\mu \nu}\partial_{\mu}X^4 \partial_{\nu} X^4\nn \\
&&\quad+\frac{3u^3}{Q_c g_s}\left(1-Q_f\left(\frac{512}{5005}+N_2\frac{4\sqrt{3}}{45}+N_3\frac{9}{20}-\fft72 \lambda_1-\lambda_2\right)u\right)\fft12\pau X^4 \pau X^4\Bigg]. \nn
\eea
The interpretation of $\lambda_1$ and $\lambda_2$ is that the correspond to coordinate transformations, and so
are actually arbitrary and one may choose these.  The numbers $N_1$ and $N_2$ are numbers that determine part of the
profile of the backreaction of the branes, and so may be constrained by some physical boundary conditions.  Here, however, we simply note that $\lambda_2$ may be chosen to eliminate the $(X^4)^2$ term completely.  This then
leaves terms of the form $(1-Q_f u C_i)$ times terms present when $Q_f=0$.  Therefore, we conclude that when
the perturbative analysis is valid, all coefficients remain the same as the $Q_f=0$ case.  Because of this, one may simply argue
that the hamiltonian of the above action is positive definite (it is a sum of squares times positive coefficients)
for the range of validity of the perturbative analysis.  Hence, we conclude that in the perturbative regime, the configuration is stable.

This depended on the leading order cancelation between the DBI and coupling to
the RR field.  Other than this, the remaining terms were all present in the $Q_f\rightarrow 0$ limit (up to gauge).  In such a case, all corrections that are order $Q_f$ cannot change the signs of coefficients,
and so stability (in the range of validity for the perturbative approach) is preserved.  We will see this again in the next section.

As a curious note, with an appropriate choice of $N_i$ and $\lambda_i$, one can completely cancel
the leading order in $Q_f$ contribution to the above action.

%%%%%%%%%%%%%%%%%%%%%%%%%%%%%%%%%%%%%%%%%%
\subsection{Stability of the Sakai Sugimoto model}
%%%%%%%%%%%%%%%%%%%%%%%%%%%%%%%%%%%%%%%%%%

Above, and in appendix C, we show that $X_4=0,X_4=\pi R_x$ is still a solution to the equations of motion resulting from the DBI+CS action.  To evaluate the second order action for fluctuations, we change
to the radial coordinate $z$ defined by $u^3=U_K^3+U_Kz^2$,
and then to the ``Cartesian'' coordinates
\be
y_1=z \sin\left(\frac{x_4}{R_x}\right) \quad \quad y_2=z \cos\left(\frac{x_4}{R_x}\right)
\ee
which also allows for comparison with the analysis performed in \cite{Sakai:2004cn}.
Here the important point is that we chose the gauge $B_1=G_1$, and so the only change
in the $z,x_4$ plane is by a conformal factor.  Hence, much of the analysis of \cite{Sakai:2004cn} follows through.
We find that the metric in these coordinates is written
\bea
&&ds^2=e^{2A_1}\left(\frac{u}{R_{D4}}\right)^{\fft32}\left(\eta_{\mu \nu}dx^{\mu} dx^{\nu}\right)+e^{2C_1}R_{D4}^{\fft32}u^{\fft12}d\Omega_4^2   \\
&& \quad + e^{2 G_1}\fft49\left(\frac{R_{D4}}{u}\right)^{\fft32}\big(\left[1-h(z)y_1^2\right]dy_1^2+\left[1-h(z)y_2^2\right]dy_2^2
-2h(z)y_1 y_2 dy_1 dy_2\big). \nn
\eea
where now all metric functions are written as functions of $y_1$ and $y_2$, and we have defined the following functions
\bea
h(z)&=&\frac{1}{z^2}\left(1-\frac{U_K}{u}\right) \nn \\
u=u(z)&=& \left(U_K^3+U_K z^2\right)^{\fft13} \\
z=z(y_1,y_2)&=& \sqrt{y_1^2+y_2^2}. \nn
\eea
Here we have suppressed the factor of $Q_f$ for ease of notation, and will only reintroduce it
at the end.
Taking the embedding $y_1(x^{\mu},u)$ one may compute the second order action the same way as the $U_K=0$ decompactification piece.  One writes the line element as $ds^2=ds_1^2+ds_2^2$ where $ds_1^2$ is diagonal and
$ds_2^2$ is already order $y_1^2$, and so again one finds that
\be
\sqrt{-g_p}=\sqrt{-g_{p0}}\left(1+\fft12 g_{p0}^{a b}h_{a b}\right)+ \mathcal{O}((y_1)^4).
\ee
One may compute the DBI action easily now,
\bea
-K_8 \int d^9\xi e^{-\phi}\sqrt{-g_p}&=& \\
&&\kern-10em -\fft23\frac{K_8 R_{D4}^{\fft32} U_K^{\fft12}V_4}{g_s}\int d^5\xi\Big(u^2e^{4A_1+4C_1+G_1-\phi_1}+\fft29\frac{R_{D4}^3}{u}e^{2A_1 +4C_1+3G_1-\phi_1}\eta^{\mu \nu}\partial_{\mu} y_1 \partial_{\nu} y_1\nn \\
&&+\fft12\frac{u^3}{U_K}e^{4A_1+4C_1+G_1-\phi_1}\left(h(z)(y_1^2-2y_1 y_2 \dot{y_1})+\dot{y_1}^2\right)\Big)\nn
\eea
where we have defined $\partial_{y_2} y_1=\dot{y_1}$.  Further, the above function $u(y_1,y_2)$ still must be expanded
in $y_1$.  The term $V_4$ is the volume of the unit four sphere, and $d^5\xi=dt dx_1 dx_2 dx_3 dy_2$.

We will integrate the term linear in $\dot{y_1}$ by parts, but first we find it convenient to introduce the
following notation
\bea
\mathcal{A}&\equiv&4A_1+4C_1+G_1-\phi_1=2A_1+2C_1+G_1-\fft12 \ph_1 \\
\mathcal{B}&\equiv&2A_1 +4C_1+3G_1-\phi_1=2C_1+3G_1-\fft12 \ph.
\eea
From the arguments in the last subsection, we expect the above combinations of fields to have the following
behavior about $y_1=0$
\bea
\mathcal{A} &=&  \mathcal{A}_0(y_2)+ \frac{2}{3}|y_1|+ \mathcal{A}_2(y_2) y_1^2+\cdots \nn \\
\mathcal{B} &=&  \mathcal{B}_0(y_2) +\cdots
\eea
where in $\mathcal{B}$ we ignore higher corrections in $y_1$ because its coefficient is already $\mathcal{O}(y_1^2)$.
In the above, we have determined the expansion in $\mathcal{A}$ of order $|y_1|$ by considering the
argument in the $u,x_4$ coordinates used to give equations (\ref{x4cusp}), and then changing coordinates to the $y_1,y_2$ variables.

In the following, we will have to evaluate $u(y_1,y_2)$ at $y_1=0$, and henceforth, we will call this function $u_y$.   Similarly we define $h_y=h(z)|_{y_1=0}$.  Plugging in the above to the second order action, and reintroducing $Q_f$ we find
\bea
-K_8 \int d^9\xi e^{-\phi}\sqrt{-g_p}&=&  \\
&&\kern-8em -\tilde{T}\int d^5\xi\Bigg[(u_y^2+y_1^2)(1+Q_f\mathcal{A}_0)+\fft29\frac{R_{D4}^3}{u_y}(1+Q_f\mathcal{B}_0)\eta^{\mu \nu}\partial_{\mu} y_1 \partial_{\nu} y_1\nn \\
&&\qquad \qquad \qquad+\fft12\frac{u_y^3}{U_K}(1+Q_f\mathcal{A}_0)\dot{y}_1^2\nn \\
&&\kern-3em+Q_f\left(u_y^2\mathcal{A}_2+\fft12\frac{u_y^3}{U_K}h_y y_2\partial_{y_2}\mathcal{A}_0\right)y_1^2
+Q_fu_y^2\fft23|y_1|\Bigg]\nn
\eea
where we define
\be
\fft23\frac{K_8 R_{D4}^{\fft32} U_K^{\fft12}V_4}{g_s}\equiv \tilde{T}
\ee
as in the work of \cite{Sakai:2004cn}.  It is easy to read off the result in \cite{Sakai:2004cn} in the $Q_f=0$ limit; it is the top two lines of the right hand side.  As in the last sections, we now add to this the contribution from $\int A_9$. This is relatively easy to do, as we find
\be
\sqrt{-g}=\fft49 R_{D4}^{\fft32}U_K^{\fft12}u_y^2\left(1+\fft23\frac{y_1^2}{\frac{u_y^3}{U_K}}+\mathcal{O}(y_1^4)\right)\sqrt{g_{\,\!_{S^4}}}
\ee
and so to second order in $y_1$ we can simply take
\be
A_{\mu_i|y_1}=\mp\frac{Q_f}{g_s}\fft49 R_{D4}^{\fft32}U_K^{\fft12}u_y^2|y_1|\left(1+\mathcal{O}(y_1^2)\right)\sqrt{g_{\,\!_{S^4}}}.
\ee
This exactly cancels the $|y_1|$ term (for the $-$ choice), as we have seen several times now (see appendix C for this occurring at the level of the equations of motion).
Therefore, the full action reads
\bea
-K_8 \int d^9\xi e^{-\phi}\sqrt{-g_p}&-&K_8\int A_9 =\nn \\
&&\kern-8em -\tilde{T}\int d^5\xi\Bigg[(u_y^2+y_1^2)(1+Q_f\mathcal{A}_0)+\fft29\frac{R_{D4}^3}{u_y}(1+Q_f\mathcal{B}_0)\eta^{\mu \nu}\partial_{\mu} y_1 \partial_{\nu} y_1 \\
&&\kern-3em+\fft12\frac{u_y^3}{U_K}(1+Q_f\mathcal{A}_0)\dot{y}_1^2
+Q_f\left(u_y^2\mathcal{A}_2+\fft12\frac{u_y^3}{U_K}h_y y_2\partial_{y_2}\mathcal{A}_0\right)y_1^2\Bigg].\nn
\eea
At this point it is sufficient to note that because all terms in the action were present before the perturbation,
we expect that wherever the perturbative analysis is valid, the stability of the Sakai Sugimoto model is maintained.  This is because whenever the perturbation is small, the above action yields a positive definite hamiltonian, and so all fluctuations will have positive energy.  Further, as we have seen in the previous section, the equations admit a perturbative solution about $u=U_K$ and so the perturbative analysis is valid from $u=U_K$ up to $u \ll \frac{1}{Q_f}$, where for the previous sections analysis gives a good approximation to the solutions.

%%%%%%%%%%%%%%%%%%%%%%%%%%%%%%%%%%%%%%%%%%%
\section{Discussion and outlook}
%%%%%%%%%%%%%%%%%%%%%%%%%%%%%%%%%%%%%%%%%%%

\begin{figure}[ht*]
%\centering
\centering
\includegraphics[width=0.5\textwidth]{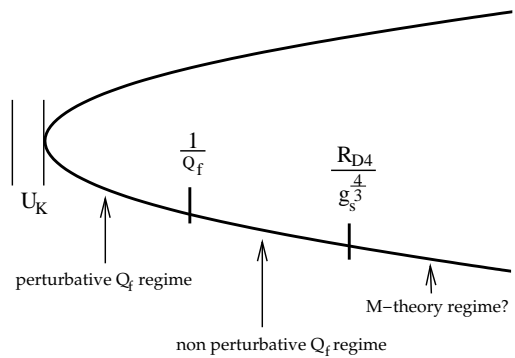}
\caption{Range of validity for calculations}
\label{hierarchy}
\end{figure}

Here we will summarize our results.  From the above calculations, we can see that when u is large enough, the solutions tend to that of the decompactified case.  The height of these functions all grow as $u$ and so to stay in the perturbative regime, we require that
\bea
&&uQ_f \ll 1 \rightarrow u \ll \frac{1}{Q_f}=\frac{4 \pi \ell_s}{g_s N_f} \rightarrow u^3 \ll \frac{64 \pi^3 \ell_s^3}{g_s^3 N_f^3}
\eea
There is a further requirement, that the supergravity approximation is valid.  This gives a restriction
\bea
g_s\left(\frac{u^3}{R_{D4}^3}\right)^{\fft14}\ll 1 \rightarrow u^3 \ll \frac{\pi N_c \ell_s^3}{g_s^3}.
\eea
One may easily compare now and see which condition is more stringent, as all coefficients of $g_s$ and $\ell_s$ are the same.  We find that generically
\be
\frac{1}{Q_f^3}< \frac{R_{D_4}^3}{g_s^4}\rightarrow 1< \frac{N_c N_f^3}{64 \pi^2}
\ee
as we assume that $N_c$ is large and $N_f \neq 0$ (see figure \ref{hierarchy}).  However, we note that for small $g_s$ the regime of validity of the perturbative backreaction (and so the validity of the probe approximation) becomes arbitrarily large.

In figure \ref{hierarchy}, we have indicated a possible M-theory (11D SUGRA) lift, although some caution is necessary,
as no know lift D8 branes is understood in the context of 11D SUGRA, at least for those described by the Romans type IIA.

We are also left with some obvious open questions:

\begin{enumerate}
\item The topic of this paper has been the low temperature limit of the Sakai Sugimoto model, and one may wish to know the qualitative differences between the low and high temperature limits.  Further, one may hope that the analysis of the high temperature limit may be easier, as the D8 branes are transverse to a cylinder, rather than a cigar.

\item It would be interesting to address the backreaction of flavor branes in other brane systems using the above techniques.  While one may worry about the perturbation breaking down near the brane for codimension other than 1, one may trust the cancelation between the DBI and CS terms in the quadratic action for fluctuations.  We believe this to be true for the following reason: for a section of brane near a smooth point in a manifold, it's backreaction (non perturbative contributions included) should behave just as the flat space case.  In such a situation, a section of parallel probe brane near by feels no force on it because the charge and mass (per unit p volume) are the same.  We may expect this to always be true.  Further, in supersymmetric situations, the supersymmetry of the backreaction may be of some assistance in fixing all coefficients.  We look forward to addressing these issues in some future work.
\end{enumerate}

\section*{Acknowledgments}
We are grateful for discussions with Oren Bergman, who was involved
in the initial stages of this project. We would like to thank Ofer
Aharony for discussions and comments on the draft of this paper.
 We also wish to thank Martha
Merzig for help editing graphs, and Josh Davis for directing us to a
useful reference. The work of B.B and J. S has been supported in
part by the Israel Science Foundation, by a grant ( DIP H.52) of
German Israel Project Cooperation  D.I.P and by the European Network
MRTN-CT-2004-512194.

This material is based upon work supported by the National Science
Foundation under Grant No. PHY-0455649.

\appendix

\section{Massive type IIA equations of Motion}
We recall the following definitions
\bea
\tilde{F}_2&&=F_2+MB_2 \nn \\
\hat{F}_4&&=F_4+\fft12M B_2\wedge B_2  \\
\tilde{F}_4&&=F_4-A_1\wedge H_3+\fft12M B_2\wedge B_2. \nn
\eea
The equations of motion for the action $S_{IIA_M}$ (\ref{MIIAaction}) are
\bea
 &&\kern-2em R_{\mu \nu}+2 \nabla_\mu \nabla_\nu \phi - \frac{1}{2\time 2!}H_\mu\,^{\rho \sigma}H_{\nu \rho \sigma}
- e^{2\phi}\frac{1}{2\cdot 2!}\left(2\tilde F_\mu\,^\rho \tilde F_{\nu \rho} -\frac{1}{2} g_{\mu \nu} \tilde F_2\cdot \tilde F_2 \right)\\
&&\kern-2em\quad \quad+e^{2\phi}\frac{1}{2\cdot 4!}\left(4\tilde F_\mu\,^{\rho_1 \rho_2 \rho_3} \tilde F_{\nu \rho_1 \rho_2 \rho_3} -\frac{1}{2} g_{\mu \nu} \tilde F_4\cdot \tilde F_4 \right)+\frac{e^{2\phi}}{2} g_{\mu \nu} M^2 =0 \nn \\
&&\kern-2emR-4\partial \phi \cdot \partial \phi -\frac{1}{2\cdot 3!}H_3\cdot H_3+4g^{\mu \nu} \nabla_\mu \nabla_\nu \phi=0 \\
&&\kern-2em \nabla_{\alpha_1}\left(e^{-2\phi}H^{\alpha_1 \beta_1 \beta_2}+A_{\alpha_2}\tilde F^{\alpha_1 \alpha_2 \beta_1 \beta_2}\right)-\frac{1}{2!\cdot 4! \cdot 4! \sqrt{-g}}\epsilon^{\beta_1 \beta_2 \cdots}F_{(4)\cdots}F_{(4)\cdots}\nn \\ && \kern-2em -M\tilde F^{\beta_1 \beta_2}
-\fft12 M B_{\alpha_1 \alpha_2}\tilde F^{\alpha_1 \alpha_2 \beta_1 \beta_2}-\frac{1}{2\cdot 2! \cdot 4!\sqrt{-g}} \epsilon^{\beta_1 \beta_2 \cdots} B_{\cdots} B_{\cdots} \hat F_{(4)\cdots}=0 \\
&&\kern-2em\nabla_{\alpha_1}\tilde F^{\alpha_1 \beta_1}-\frac{1}{3 !}\tilde F^{\beta_1 \alpha_2 \alpha_3 \alpha_4}H_{\alpha_2 \alpha_3 \alpha_4}=0 \\
&&\kern-2em\nabla_{\alpha_1}\tilde F^{\alpha_1 \beta_1 \beta_2 \beta_3}-\frac{1}{3!\cdot 4!\sqrt{-g}}\epsilon^{\beta_1 \beta_2 \beta_3\cdots}H_{\cdots} \hat F_{(4)\cdots}\nn \\
&&\kern-2em\quad \quad- \frac{M}{4!\sqrt{-g}}\epsilon^{\beta_1 \beta_2 \beta_3\cdots}
B_{\cdots}B_{\cdots}H_{\cdots}=0  \\
&&\kern-2em-\fft12 B\cdot \tilde F_2-\frac{3}{4!} B_{\cdots}B_{\cdots}\tilde F_4^{\cdots} \nn \\
&&\kern-2em\quad \quad -\frac{3}{2!\cdot 4! \cdot 4!\sqrt{-g}}
\epsilon^{\cdots}B_{\cdots}B_{\cdots}B_{\cdots}\tilde{F}_{(4)\cdots} -M +*F_{10}=0 \nn \\
&&\kern-2em dM=0
\eea
where again $\epsilon$ takes values $\pm 1$.  In the above, where we have written $\cdots$ there are indices contracted.  To reintroduce the (sub)superscripts, one puts in a set of indices in the superscripts, and then puts the same indices
in the same order in the subscripts.  In the above, we note that $M$ is constant, and may be considered piecewise constant in the presence of sources.

\section{Separating the equations}
Here we deal with the Einstein equations and the dilaton equation, and explain how to separate them.  There are 5 Einstein equations, and one for the dilaton, and we name them
\bea
&& EOM_{\ph} = 2\partial_u^2 \ph-\left(\partial_u \ph\right)^2-2\pau^2B-4\left(\partial_u A\right)^2-2\left(\partial_u B\right)^2 -4\left(\partial_u C\right)^2 \nn \\
&&\qquad +2\pa_u \ph \pa_u B -2\pa_u \ph \pa_u G+2 \pa_u B \pa_u G \nn \\
&&\qquad +e^{(2G-2B)}\Bigg(2\pax^2 \ph(u,x4)-\left(\pax \ph\right)^2-4 \left(\pax A \right)^2-2 \left(\pax G \right)^2-4 \left(\pax C\right)^2 \nn \\
&&\qquad \quad \quad \quad \quad +2\pax \ph \pax G -2\pax \ph \pax G+2\pax B \pax G -2\pax^2 G\Bigg) \label{Ep}\\
&&\qquad +12e^{2G-2C}-\frac{e^{2A-B+2G+2C+\frac{1}{2}\ph}Q_f\Delta}{g_s} \nn
\eea
\bea
&&EOM_{tt}= 4e^{(G-B+\ph)}\pau\left(e^{(-G+B-\ph)}\pau A\right) \nn \\
&&\qquad +4 e^{(2G-2B)}e^{(B-G+\ph)}\pax\left(e^{(-B+G-\ph)}\pax A\right) \label{Et} \\
&&\qquad -Q_c^2e^{4A +2G-4C+\ph}-2\frac{e^{2A-B+2G+2C+\frac{1}{2}\ph}Q_f\Delta}{g_s} \nn \\
&&EOM_{xx}= 4e^{(-B+G+\ph)}\pau\left(e^{(B-G-\ph)}\pau B\right) \nn \\
&&\qquad+4e^{(2G-2B)}\left(4(\pax A)^2+4(\pax C)^2+(\pax G)^2+e^{(B)}\pax\left(e^{(-B)}\pax(G-\ph)\right)\right) \nn \\
&&\qquad -Q_c^2e^{4A +2G-4C+\ph}+2\frac{e^{2A-B+2G+2C+\frac{1}{2}\ph}Q_f\Delta}{g_s} \label{Ex}
\eea
\bea
EOM_{uu}&=& 4\left(4(\pau A)^2+4(\pau C)^2+(\pau B)^2+e^{(G)}\pau\left(e^{(-G)}\pau(B-\ph)\right)\right) \nn \\
&&+4e^{(2G-2B)}e^{(-G+B+\ph)}\pax\left(e^{(G-B-\ph)}\pax G\right) \nn \\
&& -Q_c^2e^{4A +2G-4C+\ph}-2\frac{e^{2A-B+2G+2C+\frac{1}{2}\ph}Q_f\Delta}{g_s} \label{Eu}
\eea
\bea
EOM_{ss}&\!\!\!\!=\!\!\!\!&\!\! -4e^{(G-B+\ph)}\pau\left(e^{(-G+B-\ph)}\pau C\right) -4 e^{(2G-2B)}e^{(B-G+\ph)}\pax\left(e^{(-B+G-\ph)}\pax C\right) \nn \\
&& +12e^{(2G-2C)}-Q_c^2e^{4A +2G-4C+\ph}+2\frac{e^{2A-B+2G+2C+\frac{1}{2}\ph}Q_f\Delta}{g_s} \label{Es}
\eea
\bea
&& EOM_{m}=-4\pax A \pau A -4 \pax C \pau C+ \pax \pau \ph -\pau B \pax \ph -\pax G\pau \ph \label{Em}
\eea

Each of the above equations is to be set to zero.  The labeling we have used is that the subscript
$\ph$ denotes to the $\phi$ equation of motion, $tt$ denotes the time-time and $x_i x_i$ Einstein equations
(these are just one equation), $xx$
denotes the $x_4x_4$ equation of motion, $uu$ the $uu$, $ss$ the directions along the sphere, and $m$ the
mixed $u x_4$ equation.  Further, we have taken $\Delta=\Delta(x_4)$ to be a function only of $x_4$.

We now wish to perturb the following equations about the background solution.  For this purpose, we take
the following expansion
\bea
A(u,x_4)&=& \fft34\ln\left(\frac{3^{(\frac{1}{3})}u}{Q_c^{(\frac{1}{3})}g_s^{(\frac{1}{3})}}\right)+Q_f A_1(u,x_4)\nn\\
B(u,x_4)&=& \fft34\ln\left(\frac{3^{(\frac{1}{3})}u}{Q_c^{(\frac{1}{3})}g_s^{(\frac{1}{3})}}\right)+\fft12 \ln\left(1-\frac{U_K^3}{u^3}\right)+Q_f B_1(u,x_4)\nn \\
G(u,x_4)&=& -\fft34\ln\left(\frac{3^{(\frac{1}{3})}u}{Q_c^{(\frac{1}{3})}g_s^{(\frac{1}{3})}}\right)-\fft12 \ln\left(1-\frac{U_K^3}{u^3}\right)+Q_f G_1(u,x_4) \label{fexp} \\
C(u,x_4)&=& -\fft34\ln\left(\frac{3^{(\frac{1}{3})}u}{Q_c^{(\frac{1}{3})}g_s^{(\frac{1}{3})}}\right)+\ln(u)+Q_f C_1(u,x_4)\nn \\
\ph(u,x_4)&=& \fft32\ln\left(\frac{3^{(\frac{1}{3})}u}{Q_c^{(\frac{1}{3})}g_s^{(\frac{1}{3})}}\right)-4\ln(u)+2\ln(g_s)+Q_f\ph_1(u,x_4) \nn
\eea

It is now straightforward (and rather unilluminating) to expand the equations of motion and keep only the linear term in $Q_f$.  The only key point is that the source is already linear in $Q_f$ and so one plugs in the background fields only to the exponential $e^{2A-B+2G+2C+\frac{1}{2}\ph}$ appearing with $\Delta$. Rather than writing this out explicitly, we will simply explain the steps involved needed to separate the equations.  Henceforth when we write $EOM_{i}$ we mean the above equation of motion expanded to linear order in $Q_f$.  First, the most useful equation when expanded is equation $EOM_m$ ($\ref{Em}$), and this becomes
\be
-\frac{1}{u}\pax C_1 +\fft52 \frac{1}{u}\pax G_1-\frac{3}{u}\pax A_1 +\pax \pau \ph_1 -\fft34 \pax \ph_1\frac{(u^3+U_K^3)}{u(u^3-U_K^3)}=0.
\ee
This may be integrated to give
\be
-\frac{1}{u} C_1 +\fft52 \frac{1}{u} G_1-\frac{3}{u} A_1 + \pau \ph_1 -\fft34 \ph_1\frac{(u^3+U_K^3)}{u(u^3-U_K^3)}+F(u)=0.
\ee
One may solve this for $G_1$ and plug into the other equations.
We will denote doing so as $EOM_{i}|_{G_1}$.  One may easily solve for $F(u)$ now,
\be
2EOM_{\ph}|_{G_1}+EOM_{xx}|_{G_1}=-4\frac{\left(\pau F(u) u^4+4 F(u) u^3-U_K^3 \pau F(u) u-F(u) U_K^3\right)}{u(u^3-U_K^3)}
\ee
and so
\be
F(u)=\frac{C_F}{u(u^3-U_K^3)}.
\ee
However one can easily see that this perturbation is simply taking $U_K\rightarrow U_K+\delta U_K$ and linearizing on $\delta U_K$.  This is because under this shift, neither $\ph_1$ nor $A_1$ nor $C_1$ is changed, and so only $G_1$ shifts in $EOM_m$.  The linear shift of $G$ is given by $\frac{3 U_K^2 \delta U_k}{u^3-U_K^3}$.  Thus, we may safely absorb $F(u)$ into a shift into the definition of $U_K$.  If need be, we may always reintroduce it by shifting equations that depend on $B_1$ or $G_1$ appropriately.  Further, this is only a zero mode contribution (in $x_4$) and so will leave unaffected
much of our discussion.  For these reasons, we take $F(u)=0$ for the time being, knowing how to reintroduce it
later if need be.

At this point we have eliminated 2 equations of motion at the cost of 1 function, which puts us on
course to decouple the equations.

Next, we make the simple observation that in all equations of motion $EOM_{i}|_{G_1}$ only
$\pau B_1$ and $\pau^2 B_1$ appear.  Thus, if we can solve for $\pau B_1$, we
may eliminate $B_1$ completely.  We do so by taking $EOM_{xx}|_{G_1}-EOM_{uu}|_{G_1}$ and
solving this for $\pau B_1$.  This combination still has a delta function, and so it is important
at this step that $\Delta$ is a function only of $x_4$ so that when the expression for $\pau B_1$ is substituted
into $\pau \pau B_1$ no derivatives of delta functions appear.

We have now eliminated 3 of the 6 total equations, with the remaining combinations being
$EOM_{tt}|_{G_1,\pau B_1}$, $EOM_{xx}|_{G_1,\pau B_1}=\fft12 EOM_{\ph}|_{G_1,\pau B_1}=EOM_{uu}|_{G_1,\pau B_1}$, $EOM_{ss}|_{G_1,\pau B_1}$.  However, here we find that
\be
\fft25 EOM_{tt}|_{G_1,\pau B_1}-\fft13 EOM_{xx}|_{G_1,\pau B_1}-\frac{2}{15}EOM_{ss}|_{G_1,\pau B_1}=0
\ee
Hence, we are left with only 2 independent equations for 3 unknown functions.  This appears to be under constrained,
however, these equations are actually equations only of 2 linear combinations of the 3 functions.  The decoupled
combinations may be written
\bea
&&-3EOM_{tt}|_{G_1,\pau B_1} -\fft32 EOM_{ss}|_{G_1,\pau B_1} = \nn \\
&&\quad \quad \quad 3\pau^2F_1+\frac{3(4u^3-U_K^3)\pau F_1}{u(u^3-U_K^3)}
+\frac{g_s Q_c u^3 \pax^2 F_1}{(u^3-U_K^3)^2} \nn \\
&& \quad \quad \quad \quad \quad  -\frac{54 u F_1}{(u^3-U_K^3)}+\frac{2u(Q_c g_s)^{\fft23}\sqrt{3}}{(u^3-U_K^3)}\sqrt{\frac{u}{(Q_c g_s)^{\fft13}\left(1-\frac{U_K^3}{u^3}\right)}}\Delta \\
&&\fft 32 EOM_{tt}|_{G_1,\pau B_1}- 3 EOM_{ss}|_{G_1,\pau B_1} = \nn \\
&& \quad \quad \quad 3\pau^2F_2+\frac{3(4u^3-U_K^3)\pau F_2}{u(u^3-U_K^3)}
+\frac{ g_s Q_c u^3 \pax^2 F_2}{(u^3-U_K^3)^2} \nn \\
&& \quad \quad \quad \quad \quad  -\frac{4u(Q_c g_s)^{\fft23}\sqrt{3}}{(u^3-U_K^3)}\sqrt{\frac{u}{(Q_c g_s)^{\fft13}\left(1-\frac{U_K^3}{u^3}\right)}}\Delta
\eea
where
\bea
A_1 &=& -\fft15 F_1 + \frac{1}{10} F_2 -\frac{3}{10} \ph_1 \\
C_1 &=& \fft{1}{10} F_1 + \fft15 F_2 - \fft{1}{10} \ph_1.
\eea
We now turn to the question of fixing $\ph$.  For this purpose, we remember that we used the
combination
\bea
&& EOM_{xx}-EOM_{uu}= \nn \\
&& \quad \quad  4 \pau^2 \ph_1 -\fft43\frac{u^3Q_cg_s \pax^2 \ph_1}{\left(u^3-U_K^3\right)^2}-\frac{2\left(4\pau C_1 -5 \pau G_1 -5 \pau B_1 +12 \pau A_1\right)}{u} \nn \\
&& \quad \quad+\frac{\fft43 u(Q_c g_s)^{\fft23}\sqrt{3}}{(u^3-U_K^3)}\sqrt{\frac{u}{(Q_c g_s)^{\fft13}\left(1-\frac{U_K^3}{u^3}\right)}}\Delta
\eea
to solve for $\pau B_1$.  It is clear for our setup that $\ph$ must have some ``kink'' part in its solution
to account for the delta function, as the only $x_4$ derivatives that appear act on $\ph$.  However, by adding
a zero $-8/u\pau(u \int{EOM_{m}dx_4})$ to the above expression, we find
\bea
&& EOM_{xx}-EOM_{uu}-\fft8u\pau(u \int{EOM_{m},dx_4})= \nn \\
&& \quad \quad  -4 \pau^2 \ph_1 -\fft43\frac{u^3Q_cg_s \pax^2 \ph_1}{\left(u^3-U_K^3\right)^2}-\frac{2\left(u^3-7U_K^3\right)\pau \ph_1}{u\left(u^3-U_K^3\right)}-\frac{36 u U_K^3 \ph_1}{\left(u^3-U_K^3\right)^2} \nn \\
&& \quad \quad+\frac{\fft43 u(Q_c g_s)^{\fft23}\sqrt{3}}{(u^3-U_K^3)}\sqrt{\frac{u}{(Q_c g_s)^{\fft13}\left(1-\frac{U_K^3}{u^3}\right)}}\Delta +\frac{10}{u}\left(\pau B_1-\pau G_1\right)
\eea
Now it becomes clear how one may maintain continuity of the functions and at the same time
separate the equations.  We take $B_1=G_1$
and then solve the remaining equation above.
One may have guessed this gauge, as one can bring any two dimensional metric to a conformally
flat one.  We do not impose this on the full metric, however, as the polynomials in $u$
are easier to work with.

One more comment is in order.  If one wishes, one may linearize on a small change
$U_K \rightarrow U_K+\delta U_K$.  Under this, $B_1$ and $G_1$ transform differently.  This
can give a new source term to the equation for $\ph_1$.  However, this change only affects the
zero mode (in $x_4$) of $\ph_1$, and hence will not affect the shape of $\ph_1$ in the $x_4$
direction.
%%%%%%%%%%%%%%%%%%%%%%%%%%%%%%%%%%%%%%%%%%%%%%%%%%%%%%%%%%%%%%%
\section{EOM for DBI+CS: details of $x^4=0$ solution}
%%%%%%%%%%%%%%%%%%%%%%%%%%%%%%%%%%%%%%%%%%%%%%%%%%%%%%%%%%%%%%%

Here we find the equations of motion for the embedding functions $X^{\mu}$ resulting from the action
\be
-\frac{g_s S_B}{Q_f}=\int d^{9}\xi e^{-\phi} \sqrt{-g_p} + \int A_{9},
\ee
and explicitly show that $x_4=0$ ($\pi R_x$) is still a solution.
We will find the equations of motion for the first part, and then turn our attention to the second part
of the above action.  First, we change frame by scaling the metric $g_{string}=\exp{\left(\fft29\phi\right)} G$ to write
the first part of the action
\be
S_D \equiv \int d^{9}\xi \sqrt{-G_p}
\ee
The equations of motion for the fields $X^\mu(\xi)$ in the above action are
\be
\frac{\delta S_D}{\delta X^\mu}=-\sqrt{-G_p} G_{\mu \nu}\left(\nabla_p^2 X^\nu+G_p^{a b}\frac{\partial X^\alpha}{\partial \xi^a} \frac{\partial X^\beta}{\partial \xi^b}\Gamma^\nu_{\alpha \beta}\right)
\ee
where objects with a $p$ subscript are constructed using the pullback metric, and $\Gamma$ is the full
spacetime Christoffel connection.  We wish to ask whether $X^i=\xi^i$ for $i\neq x_4$ and $X^4=x_4=$constant
is a solution to the equations of motion.  Consider first the $X^k$:
\bea
\frac{\delta S_D}{\delta X^k}&&=-\sqrt{-G_p} G_{ki}\Bigg(\frac{1}{\sqrt{-G_p}}\partial_a G_p^{a b} \sqrt{-G_p} \partial_b X^i  \\
&&\kern 8em +G_p^{a b}\partial_a X^\alpha  \partial_b X^\beta \fft12G^{i\rho}\left(G_{\rho \alpha,\beta}+G_{\rho \beta,\alpha}-G_{\alpha \beta,\rho}\right)\Bigg) \nn
\eea
where we use $\partial_a$ as shorthand for a partial derivative in $\xi^a$, and we have used the fact
that our metric is diagonal.  Choosing the $X^i=\xi^i$ removes the derivative from the first part of the
equation.  Also, the fact that $G$ is diagonal, and identical to the pullback metric for indices
$i,j$, allows us to simplify the above further
\bea
&&=-\sqrt{-G_p} G_{(p) ki}\Bigg(\frac{1}{\sqrt{-G_p}}\partial_i G_p^{i i} \sqrt{-G_p} \\
&&\kern 8em +G_p^{a b}\partial_a X^\alpha  \partial_b X^\beta\fft12G_p^{ii}\left(\delta_{\alpha i}G_{ii,N}+\delta_{\beta i}G_{ii,M}-G_{\alpha \beta,i}\right)\Bigg)\nn
\eea
where $i$ inside the parentheses are not summed.  Note that the $\delta_{Ni}$ projects the last remaining
metric down to the pullback metric, and we are left with
\be
=-\sqrt{-G_p} G_{(p) ki}\left(\frac{1}{\sqrt{-G_p}}\partial_i G_p^{i i} \sqrt{G_p} +G_p^{ii} G_p^{ii} G_{ii,i}- \fft12 G_p^{ii} G_p^{ab}G_{ab,i}\right)
\ee
which simplifies further to
\be
=-\sqrt{-G_p} G_{(p) ki}\left(G^{i i}_{(p),i} +\fft12 G_p^{ii}G_p^{a b}G_{(p)a b,i} +G_p^{ii} G_p^{ii} G_{(p)ii,i}- \fft12 G_p^{ii} G_p^{ab}G_{ab,i}\right).
\ee
This is obviously zero: the metric is diagonal, so that
\be
G_p^{ii}G_{(p)ii}=1 \rightarrow G_{(p),i}^{ii}G_{(p)ii}+G_p^{ii}G_{(p)ii,i}=0
\ee
which then causes the first and third terms to cancel.

Hence, we are left with evaluating the $X^4$ equation of motion:
\be
-\sqrt{-G_p} G_{44}\left(\nabla_p^2 X^4+G_p^{a b} \partial_a X^\mu \partial_b X^\nu \Gamma^4_{\mu \nu}\right)
\ee
where we use the shorthand $4$ to mean the $x_4$ components.

The first term vanishes as $X_4=$constant.  The second term we evaluate similarly to the
last discussion
\bea
&&-\sqrt{-G_p} G_{44}\left(\nabla_p^2 X^4+G_p^{a b} \partial_a X^L \partial_b X^M \Gamma^4_{L M}\right) \nn \\
&=&-\sqrt{-G_p} G_{44}\left(G_p^{a b}\partial_a X^\mu \partial_b X^\nu \fft12G^{44}\left(\delta_{\mu 4} G_{44,\nu}+\delta_{\nu 4} G_{44,\mu}-G_{\mu \nu,4}\right)\right).
\eea
This time, however, the $\delta_{M 4}$ and $\delta_{N 4}$ give zero (as the $X^4=$constant), and hence only the
last term remains
\bea
&=&\sqrt{-G_p} \fft12 G_{44}G^{44}G_p^{a b}G_{(p)ab,4}\nn \\
&=&\sqrt{-G_p}\frac{1}{\sqrt{-G_p}} \partial_{x_4}\sqrt{-G_p} \\
&=&\sqrt{-G_p}\partial_{x_4}\left(4A_G+G_G+4C_G\right) \nn
\eea
where the subscripts are to denote the new frame that we switched to.  Switching back to the
string frame metric, we find
\be
\sqrt{-g_p}e^{-\phi}\partial_{x_4}\left(-\fft12\ph+G+2A+2C\right).
\ee
At this point we stop this analysis because we must also look at the equations of motion coming from the CS action, as this has a cusp as well.

First, for the embeddings that we have chosen, the background form field is the following
\bea
F_{10}=*M=*\frac{Q_f}{g_s}&&\quad A_9=\Omega_9\frac{Q_f}{g_s}e^{B}|x_4| \nn \\
&&A_{012356789}=\sqrt{-g_p}\frac{Q_f}{g_s}e^{B}|x_4|
\eea
where $x_4\in \{-\pi R_x \cdots \pi R_x\}$, producing both the positive and negative delta function.  In the
above, we must use only zeroth order metric functions, as the above statement is already linear in
$Q_f$.  The function $g_p$ is the determinant of the metric setting $x_4=$constant, i.e. the pullback metric
on the D8s.

Next, we wish to consider the action
\be
S_A\equiv \int_{D8} A_9=\frac{1}{9!}\int d^9\xi A_{M_1\cdots M_9} \partial_{a_1} X^{M_1}\cdots \partial_{a_9} X^{M_9}\epsilon^{a_1\cdots a_9}
\ee
where the epsilon takes values $\pm1$.  This is very easy to vary w.r.t. the fields $X^I$:
\bea
\frac{\delta S_A}{X^I}&=&-\frac{9}{9!}\partial_{a_1}A_{I M_2\cdots M_9} \partial_{a_2} X^{M_2}\cdots \partial_{a_9} X^{M_9}\epsilon^{a_1\cdots a_9}\nn \\
&&\quad +\frac{1}{9!}\partial_I(A_{M_1\cdots M_9}) \partial_{a_1} X^{M_1}\cdots \partial_{a_9} X^{M_9}\epsilon^{a_1\cdots a_9}\nn \\
\eea
Taking $I=i$ to be one of the directions along the world volume coordinates, we see that the first term
and second term are identical.  This is because in the first equation $i$ and $a_1$ must agree to give a non
zero answer (when contracting the epsilon).  Of course in the sum there are $8!$ occurences of this.  Hence,
the $9!$ cancels, and one simply gets $\partial_i A_{0...9|x_4}$, where we use $|x_4$ to indicate that the
index for $x_4$ has been omitted.  The second term is also equal to this, as the contraction yields a ($9!$).
Hence the full equations of motion for the $X^I$ fields along the volume coordinates vanish.

Next, taking the above equation of motion for the $x_4$ direction, one finds only a contribution from the second
part, i.e.
\be
\frac{\delta S_A}{X^{x_4}}=\partial_{x_4}(A_{0\cdots 9|x_4}) \\
\ee
Therefore, the full equation of motion for the field $X^4$ assuming that
it is constant reads
\be
-\left(\sqrt{-g_p}e^{-\phi}\partial_{x_4}\left(-\fft12\ph+G+2A+2C\right)\pm\partial_{x_4}(A_{0\cdots 9|x_4})\right)
\ee
where we have restored a $-$ sign earlier omitted in front of the DBI action.  Also, the $\pm$ is to be read
whether we are putting $D8$ or $\bar{D8}$ branes in the background.  Both
are linear order in $Q_f$ because to zeroth order none of the metric components depend
on $x_4$, and $A_9$ is already linear in $Q_f$.  Hence, all other metric components are
set to being their background values (except those with the $x_4$ derivative).
Considering the cusp solution near $x_4=0$ one reads (factoring out $Q_f$)
\bea
&& -\sqrt{-g_p}\frac{u^{-\fft34}}{gs R_{D4}^{-\fft34}}\partial_{x_4}\left(
\frac{u^{\fft32}\left(1-\frac{U_K^3}{u^3}\right)^{\fft12}|x_4|}{(R_{D4})^{\fft32}}+\mathcal{O}(x_4^2)\right) \nn \\
&& \quad-\pm\partial_{x_4}\left(
\sqrt{-g_p}|x_4|\frac{1}{g_s} \frac{u^{\fft34}\left(1-\frac{U_K^3}{u^3}\right)^{\fft12}}{R_{D4}^{\fft34}}+\mathcal{O}(x_4^2)\right) \nn \\
&=&-\sqrt{-g_p}\frac{u^{\fft34}\left(1-\frac{U_K^3}{u^3}\right)^{\fft12}}{R_{D4}^{\fft34}g_s}\partial_{x_4}\left(|x_4|\pm |x_4|\right)
\eea
where in the third line, we ignore the higher order in $x_4$ corrections, as we will evaluate the
derivative at $x_4=0$.  Above we have also switched back to the more familiar $g_s Q_c=3R_{D4}^3$ notation.

\end{document}